\documentclass[12pt]{article}
\usepackage{fontspec}
\usepackage{amsmath}
\usepackage{amsfonts}
\usepackage{unicode-math}
\usepackage{sourceserifpro}
\usepackage{microtype}
\usepackage{ragged2e}
\usepackage{parskip}
\usepackage{amsthm}
\usepackage{rotating}
\usepackage{threeparttable}
\usepackage{array}
\usepackage{graphicx}
\usepackage{caption}
\usepackage{multirow}
\usepackage{booktabs}
\usepackage{setspace}
\usepackage{natbib}
\usepackage{hyperref}
\usepackage{siunitx}

\hypersetup{
  colorlinks=true,
  linkcolor=red,
  citecolor=red,
  urlcolor=red,
  pdfborder={0 0 0}
}
\usepackage[export]{adjustbox}
\usepackage{csquotes}
\bibpunct{(}{)}{;}{a}{,}{,}
\usepackage{enumerate}

\newtheorem{proposition}{Proposition}
\newtheorem{corollary}{Corollary}[proposition]

\newtheorem{definition}{Definition}

\newenvironment{keywords}{\par\vspace{0.5em}\noindent\textbf{Keywords:} }{\par}
\newenvironment{jel}{\par\vspace{0.25em}\noindent\textbf{JEL Classification:} }{\par}

\newcommand{\om}{\omega}

\title{\textbf{Credit Capacity and the Propagation of Funding Shocks: Evidence from U.S. and Brazilian Financial Intermediaries}}

\author{
  Ayush Jha\footnote{Corresponding Author. Department of Economics, Texas Tech University,
    Lubbock, TX 79409; \texttt{ayush.jha@ttu.edu}}
  \quad
  Ali Jaffri\footnote{College of Business, North Dakota State University,
    Fargo, ND 58105; \texttt{Ali.Jaffri@ndsu.edu}}
  \quad
  Frank J.\ Fabozzi\footnote{Carey Business School, Johns Hopkins University,
    Baltimore, MD 21218; \texttt{ffabozz1@jhu.edu}}
}
\date{}

\begin{document}
\maketitle

\begin{abstract}
Why do similar funding shocks generate sharply different credit outcomes across countries? We develop and estimate a dynamic structural model in which intermediary credit capacity governs the transmission of funding disruptions to lending. Using supervisory data on U.S. banks and credit unions and Brazilian banks and cooperatives from 2002--2025, we recover institution-level credit capacity and its dynamics across major crisis episodes. Credit capacity is three to six times larger in the United States than in Brazil, while persistence is similar across countries. As a result, funding shocks generate substantially larger and more persistent lending contractions in Brazil. Counterfactual analysis shows that differences in baseline credit capacity, rather than persistence, account for most cross-country variation in crisis propagation and policy effectiveness.
\end{abstract}

\begin{keywords}
Financial intermediation; Funding shocks; Institutional depth; Credit capacity;
Crisis propagation; Impulse response; Policy counterfactuals.
\end{keywords}

\begin{jel}
G21; E44; O16; G28; C61.
\end{jel}

\onehalfspacing

\section{Introduction}
\label{sec:intro}

Why do some funding shocks evolve into prolonged credit contractions while others generate only temporary disruptions? The answer is often attributed to differences in the size of the shock, the capitalization of financial institutions, or the aggressiveness of policy intervention. Yet these explanations do not fully account for why seemingly similar shocks can produce markedly different levels of financial resilience across banking systems. This paper argues that an additional factor is central: the capacity of financial intermediaries to transform wholesale funding into credit supply. Economies with greater intermediary capacity are more resilient to funding disruptions and can maintain credit provision during periods of stress, whereas economies with limited capacity experience persistent contractions in lending even when facing similar shocks. 

We study this mechanism through the interaction between commercial banks and mutual intermediaries. In both the United States and Brazil, commercial banks provide wholesale liquidity to networks of credit unions and cooperatives whose lending activity depends on access to external funding. When funding conditions deteriorate, the key determinant of lending outcomes is not simply the availability of wholesale funds but the extent to which those funds can be converted into credit. We refer to this conversion factor as intermediary lending capacity.

To formalize this idea, we develop a dynamic structural model in which lending capacity is an endogenous state variable governing the transmission of funding shocks to credit supply. The model embeds a standard lending constraint within a recursive framework and allows intermediary capacity to evolve over time in response to capitalization, crisis conditions, and policy intervention. This structure generates a simple prediction: the resilience of a financial system during periods of stress depends not only on the persistence of funding shocks but also on the level of lending capacity available before the shock occurs. Consequently, financial stability is shaped not only by the shocks that institutions face but also by the capacity buffers embedded within the intermediary sector prior to the onset of stress.

We estimate the model using quarterly supervisory data for banks, credit unions, and cooperative financial institutions in the United States and Brazil from 2002 to 2025. The analysis combines structural recovery of institution-level lending capacity with dynamic panel estimation, local-projection impulse responses, and counterfactual policy experiments. The cross-country setting is particularly useful because the two financial systems share broadly similar intermediary structures while differing substantially in the depth of their funding networks.

Three findings emerge regarding the determinants of financial resilience and crisis propagation. First, lending capacity is substantially higher in the United States than in Brazil. Across crisis episodes, estimated steady-state capacity in the United States exceeds its Brazilian counterpart by a factor of approximately three to six. Second, capacity dynamics are highly persistent in both countries, implying similar rates of recovery from funding disruptions. Third, despite comparable persistence, lending outcomes differ dramatically across countries because institutions operate with very different levels of baseline capacity. Funding shocks generate relatively short-lived lending effects in the United States but prolonged credit contractions in Brazil.

The counterfactual analysis further shows that interventions designed to expand intermediary capacity are most effective when a large fraction of institutions operate near their lending constraint. These results suggest that intermediary capacity may serve as a useful state variable for evaluating systemic vulnerability and the likely effectiveness of stabilization policies. Consequently, identical policy interventions produce substantially larger lending effects in environments characterized by limited baseline capacity. By contrast, differences in the persistence of capacity shocks explain relatively little of the observed cross-country variation in lending outcomes.

This paper contributes to the literature in three ways. First, it introduces a structural measure of intermediary lending capacity that can be recovered at the institution level and tracked over time. Second, it develops a dynamic framework linking funding conditions to lending outcomes through the evolution of that capacity. Third, it provides evidence that differences in steady-state lending capacity are a primary determinant of cross-country variation in financial resilience, crisis propagation, and the ability of financial systems to sustain credit provision during periods of stress.

More broadly, the findings suggest that the architecture of intermediary funding relationships plays a central role in financial stability. Existing work has emphasized capital adequacy, liquidity management, and policy intervention as determinants of crisis outcomes. The evidence presented here highlights an additional margin: the capacity with which wholesale funding can be transformed into credit supply. Intermediary lending capacity functions as a resilience mechanism that influences the extent to which funding shocks are absorbed or amplified within the financial system. Differences in this capacity help explain why some financial systems recover quickly from stress events while others experience prolonged disruptions in credit provision. Understanding how that capacity is accumulated and sustained may therefore be essential for explaining why otherwise similar funding shocks produce markedly different lending outcomes across financial systems.

The results imply that measures of intermediary lending capacity may provide useful information regarding systemic vulnerability, the likely severity of credit contractions, and the effectiveness of stabilization policies during periods of financial stress.

The remainder of the paper proceeds as follows. Section~\ref{sec:lit} discusses the related literature. Section~\ref{sec:model} presents the model and identification strategy. Section~\ref{sec:data} describes the data and variable construction. Section~\ref{sec:results} presents the empirical findings and counterfactual analysis. Section~\ref{sec:conclusion} concludes.

\section{Related Literature}
\label{sec:lit}

This paper contributes to four strands of the banking and financial intermediation literature.

The first strand studies the transmission of funding shocks to credit supply. A large empirical literature documents that disruptions to bank funding conditions affect lending outcomes through both price and quantity margins. Early evidence in \citet{kashyap2000million} and \citet{khwaja2008tracing} establishes the existence of a bank-lending channel, while \citet{jimenez2012credit} show that capital and liquidity positions shape the sensitivity of lending to credit-supply shocks. Subsequent work emphasizes the importance of funding structure, including the role of deposits \citep{DrechslerSavovSchnabl2017} and wholesale funding markets \citep{ivashina2010bank}. This literature identifies the existence of funding transmission channels but generally treats intermediary lending capacity as either fixed or implicit. We instead recover a time-varying lending-capacity multiplier from a structural model and estimate its dynamics directly.

The second strand examines balance-sheet constraints, liquidity creation, and intermediary behavior. Foundational contributions by \citet{diamond1983bank} and \citet{DiamondRajan2001} characterize banks as liquidity-transforming institutions whose funding structure is central to credit creation. Related work shows that intermediary capital constraints affect both investment and lending decisions \citep{10.1162/003355397555316}, while \citet{berger2009bank} document substantial heterogeneity in liquidity creation across institutions. \citet{AcharyaShinYorulmazer2011} further demonstrate that liquidity management and crisis interventions can influence the severity of funding contractions. Our framework builds on this literature by introducing an endogenous state variable that links balance-sheet conditions to lending capacity over time and generates testable predictions regarding the persistence of crisis effects.

The third strand studies crisis transmission and financial resilience across institutional environments. \citet{LaevenValencia} show that policy responses play an important role in determining the duration and cost of banking crises, while \citet{EveraertPozzi} document persistent effects of financial crises on measures of financial development across countries. Related evidence in \citet{Sever} suggests that institutional quality influences both crisis severity and recovery. We contribute to this literature by identifying a micro-founded transmission mechanism through which institutional structure affects the propagation of funding shocks. In our framework, differences in crisis outcomes arise not only from the shocks themselves but also from differences in steady-state lending capacity and its dynamic evolution.

The fourth strand emphasizes funding networks and nonbank financial intermediation. Research on money market funds, bond funds, and other market-based intermediaries documents how funding disruptions can propagate through interconnected financial institutions \citep{kacperczyk2013safe,10.1093/rfs/hhu025,goldstein2017investor}. Related work on credit lines and liquidity insurance highlights the importance of contingent funding arrangements for sustaining credit supply during periods of stress \citep{gatev2006banks,acharya2013aggregate,campello2010real}. At a broader level, \citet{acemoglu2015systemic} show that network structure governs whether local shocks are amplified or absorbed. Our analysis complements this perspective by introducing an institution-level measure of network capacity. The steady-state multiplier, $\lambda^{SS}$, captures the extent to which funding networks can absorb shocks before lending constraints become binding.

Our contribution is threefold. First, we identify a time-varying lending-capacity multiplier from structural equilibrium conditions rather than treating intermediary capacity as exogenous. Second, we estimate the dynamic evolution of that multiplier and show that the persistence of capacity shocks is remarkably similar across countries. Third, we provide a structural cross-country analysis demonstrating that differences in lending outcomes following crises are driven primarily by differences in steady-state lending capacity rather than differences in shock persistence. Together, these results introduce intermediary lending capacity as a measurable and economically important determinant of financial resilience.

\section{Model and Identification}
\label{sec:model}

\subsection{Economic Environment}

Time is discrete. A continuum of mutual intermediaries indexed by $i\in[0,1]$ operate
in local credit markets. At the beginning of period $t$, the aggregate state
\[
  Z_t = (C_t,\,\Xi_t,\,i_t,\,\mathrm{Risk}_t)
\]
is realized, where $C_t\in\{0,1\}$ denotes a crisis state, $\Xi_t$ captures the
intensity of liquidity-support programs, $i_t$ is the policy rate, and
$\mathrm{Risk}_t$ is an aggregate credit-risk index. Each intermediary observes its
balance-sheet position, consisting of equity capital $K_{it}$, deposits $D_{it}$, and
committed bank credit lines $BL_{it}$, together with its lending-capacity parameter
$\lambda_{it}$ and funding cost $\omega_{it}$. Lending decisions are then made,
profits are realized, and state variables evolve according to the laws of motion
specified below.

A key feature of the environment is that deposits and committed credit lines are
predetermined within the period. Consequently, lending capacity is constrained by the
intermediary's existing funding base rather than contemporaneous financing decisions.

\subsection{Static Equilibrium}

Loan demand in market $j$ is given by
\begin{equation}
  L^{D}_{jt} = A_{jt} - \chi r^{L}_{it} + \zeta Z_{jt},
  \qquad \chi>0,
  \label{eq:demand}
\end{equation}
where $A_{jt}$ summarizes local demand conditions and $\chi$ governs the interest
elasticity of borrowing.

The intermediary's marginal funding cost is
\begin{equation}
  \omega_{it}
  = \theta_1 r^{D}_t
  + \theta_2 r^{BL}_t
  + \theta_3 \kappa_t
  + \eta_{it},
  \label{eq:omega}
\end{equation}
where $r^{D}_t$ denotes the deposit rate, $r^{BL}_t$ the wholesale borrowing rate,
and $\kappa_t$ a regulatory-capital index. Wholesale funding costs satisfy
\[
  r^{BL}_t = i_t + \gamma_0 + \gamma_1\mathrm{Risk}_t - \gamma_2\Xi_t + \varepsilon_t,
\]
so liquidity-support programs reduce the cost of external funding.

Loan supply is characterized by
\begin{equation}
  r^{L}_{it} = \omega_{it} + \phi L_{it} + \mu_{it},
  \qquad \phi>0, \quad \mu_{it}\ge 0,
  \label{eq:supply}
\end{equation}
where $\phi$ captures convex intermediation costs and $\mu_{it}$ is the multiplier
associated with the balance-sheet constraint
\begin{equation}
  L_{it} \;\le\; D_{it} + \lambda_{it}BL_{it},
  \qquad
  \mu_{it}\!\left(L_{it} - D_{it} - \lambda_{it}BL_{it}\right) = 0.
  \label{eq:constraint}
\end{equation}
The parameter $\lambda_{it}$ measures the efficiency with which committed credit lines
are transformed into lending capacity. Larger values of $\lambda_{it}$ correspond to
deeper funding relationships and greater effective access to wholesale liquidity.

Combining demand and supply yields equilibrium lending
\begin{equation}
  L^{*}_{it}
  = \min\!\left\{L^{U}_{it},\; D_{it}+\lambda_{it}BL_{it}\right\},
  \label{eq:Lstar}
\end{equation}
where unconstrained lending is
\begin{equation}
  L^{U}_{it}
  = \frac{A_{jt} - \chi\omega_{it} + \zeta Z_{jt}}{1+\chi\phi}.
\end{equation}
When the constraint is slack, lending equals the unconstrained equilibrium quantity.
When the constraint binds, lending is determined entirely by balance-sheet capacity.
The associated shadow value is
\begin{equation}
  \mu^{*}_{it}
  = \frac{1+\chi\phi}{\chi}
    \left(L^{U}_{it} - D_{it} - \lambda_{it}BL_{it}\right),
  \qquad \mu^{*}_{it}>0.
  \label{eq:mu}
\end{equation}
The shadow value summarizes the marginal increase in lending that would result from an
incremental relaxation of funding capacity.

\subsection{Dynamic Capacity Formation}

The static framework identifies contemporaneous funding constraints but does not
characterize their persistence. To study the propagation of crises and the dynamic
effects of policy interventions, lending capacity is modeled as an endogenous state
variable.

The economic motivation is that credit-line relationships between intermediaries and
commercial banks adjust gradually. Funding disruptions therefore affect lending
capacity beyond the period in which the shock occurs. Three parameters govern this
process. The persistence parameter $\rho\in(0,1)$ captures the durability of funding
relationships. The capital-buffer parameter $\psi\ge 0$ measures the extent to which
intermediary capitalization supports future funding access. The policy elasticity
parameter $\gamma_{\Xi}\ge 0$ captures the effect of liquidity-support programs on
aggregate lending capacity.

\subsection{Dynamic Optimization}

Per-period profits are
\begin{equation}
  \Pi_{it}(L_{it})
  = a_{it}L_{it} - \tfrac{b}{2}L_{it}^{2},
  \label{eq:profit}
\end{equation}
where
\[
  a_{it}
  = \frac{A_{jt}+\zeta Z_t}{\chi} - \omega_{it},
  \qquad
  b = \frac{1+\chi\phi}{\chi}.
\]
Capital evolves according to
\begin{equation}
  K_{i,t+1} = (1-\delta_K)K_{it} + \Pi^{*}_{it} - F,
  \label{eq:K_law}
\end{equation}
where $\delta_K$ captures payouts and write-downs and $F$ is a fixed operating cost.

The intermediary solves
\begin{equation}
  V(s_{it})
  = \max_{0\,\le\, L\,\le\, D_{it}+\lambda_{it}BL_{it}}
    \left\{
      \Pi_{it}(L)
      + \beta\, E_t\!\left[V(s_{i,t+1})\right]
    \right\},
  \label{eq:bellman}
\end{equation}
with state vector
\[
  s_{it} = (K_{it},\,D_{it},\,BL_{it},\,\lambda_{it},\,\omega_{it},\,Z_t).
\]
Applying the envelope condition yields the marginal value of lending capacity,
\begin{equation}
  p_t
  \;\equiv\;
  \frac{\partial V}{\partial \lambda_{it}}
  = \mu^{*}_{it}BL_{it}
  + \beta\rho\, E_t[p_{t+1}],
  \label{eq:shadow}
\end{equation}
which admits the forward representation
\begin{equation}
  p_t
  = \sum_{h=0}^{\infty}
    (\beta\rho)^{h}\,
    E_t\!\left[\mu^{*}_{i,t+h}\,BL_{i,t+h}\right].
  \label{eq:shadow_forward}
\end{equation}
The value of funding capacity therefore reflects both current lending opportunities
and future periods in which balance-sheet constraints remain operative.

\subsection{The Law of Motion for Capacity}

The evolution of lending capacity is governed by
\begin{equation}
  \lambda_{i,t+1}
  = \rho\lambda_{it}
  + (1-\rho)\left[\bar{\lambda}_t + \psi k_{it}\right]
  + \sigma_{\lambda}\xi_{it},
  \label{eq:lom}
\end{equation}
where
\[
  \bar{\lambda}_t = \bar{\lambda} + \alpha_{\lambda}C_t + \gamma_{\Xi}\Xi_t
\]
is the aggregate capacity target and $k_{it}\equiv K_{it}/A_{it}$ denotes the
capital-to-asset ratio.

Equation~\eqref{eq:lom} decomposes capacity dynamics into persistence, mean
reversion, and idiosyncratic shocks. The parameter $\rho\in(0,1)$ captures the
persistence of intermediary funding relationships. High values of $\rho$ correspond
to durable credit-line arrangements and slow adjustment of wholesale funding capacity.
The term $(1-\rho)(\bar{\lambda}_t+\psi k_{it})$ governs convergence toward a
time-varying target. Aggregate conditions shift the target through crisis effects
($\alpha_{\lambda}<0$) and policy support ($\gamma_{\Xi}>0$), while intermediary
capitalization affects long-run funding access through the capital premium $\psi$. The
innovation $\sigma_{\lambda}\xi_{it}$ captures institution-specific changes in funding
relationships that are orthogonal to aggregate conditions.

In the absence of crisis shocks, the stationary distribution of lending capacity is
\begin{equation}
  \lambda_i^{SS}
  \sim \mathcal{N}\!\left(
    \bar{\lambda} + \gamma_{\Xi}\Xi_0 + \psi k^{SS},\;
    \frac{\sigma_{\lambda}^{2}}{1-\rho^{2}}
  \right).
  \label{eq:ss}
\end{equation}
The steady-state mean determines the average lending capacity generated by a unit of
committed funding, while the cross-sectional variance is increasing in $\rho$.
Consequently, systems characterized by more persistent funding relationships exhibit
greater heterogeneity in lending capacity across institutions.

Equations~\eqref{eq:K_law}, \eqref{eq:lom}, and~\eqref{eq:shadow_forward} jointly
imply a dynamic complementarity between capitalization and lending capacity. Higher
capital increases future funding access through $\psi$, which raises the continuation
value of capacity and strengthens incentives for balance-sheet accumulation.

\subsection{Crisis Dynamics}

Consider a crisis shock $C_t=1$ occurring at $t=0$ from the steady state characterized
by~\eqref{eq:ss}. The expected path of lending capacity is
\begin{equation}
  \mathbb{E}_0[\lambda_{i,h}]
  = \lambda^{SS}
  + \rho^{h}\alpha_{\lambda}
  + (1-\rho^{h})\gamma_{\Xi}\,
    \mathbb{E}_0\!\left[
      \sum_{j=1}^{h} \rho^{h-j}\left(\Xi_j - \Xi_0\right)
    \right].
  \label{eq:crisis_path}
\end{equation}
The crisis effect decays geometrically at rate $\rho$, while policy interventions
shift the transition path through their effect on the aggregate capacity target. In the
absence of policy support, the impulse response of lending capacity simplifies to
\begin{equation}
  \mathrm{IRF}_{\lambda}(h) = \rho^{h}\alpha_{\lambda},
  \label{eq:irf}
\end{equation}
with half-life
\begin{equation}
  h_{1/2} = \frac{\log 2}{\log(1/\rho)}.
\end{equation}
Equation~\eqref{eq:crisis_path} highlights three distinct determinants of post-crisis
outcomes: the initial contraction $\alpha_{\lambda}$, the persistence parameter $\rho$,
and the pre-crisis level of lending capacity $\lambda^{SS}$. These components affect,
respectively, the magnitude, duration, and baseline resilience of the lending system.
Policy support enters cumulatively through the persistence channel, implying larger
effects when capacity distortions are highly persistent.

\subsection{Empirical Predictions}
\label{sec:predictions}

The model yields three predictions.

\paragraph{Prediction 1: Steady-state capacity is the primary determinant of lending
resilience.}
For a given persistence parameter $\rho$, systems with higher $\lambda^{SS}$ experience
smaller lending contractions following an identical capacity shock because fewer
intermediaries become capacity constrained.

\paragraph{Prediction 2: Policy effectiveness is increasing in the prevalence of binding
constraints.}
The cumulative policy multiplier is
\[
  M^{\Xi}(h) = \gamma_{\Xi} \sum_{j=0}^{h} \rho^{h-j}.
\]
The quantity effects of policy interventions are largest when a substantial fraction of
intermediaries operate at the lending-capacity constraint.

\paragraph{Prediction 3: Persistent funding relationships generate slower adjustment but
higher long-run capacity.}
Higher values of $\rho$ imply slower recovery from adverse shocks while supporting
larger steady-state lending capacity through more durable funding relationships.

\subsection{Identification}
\label{sec:identification}

Estimation proceeds in three stages.

\paragraph{Stage 1: Recovery of lending capacity.}
For observations satisfying $\hat{\mu}_{it}>0$, lending capacity is identified from the
binding constraint as
\begin{equation}
  \hat{\lambda}_{it}
  = \frac{L_{it} - D_{it}}{BL_{it}}.
  \label{eq:lam_rec}
\end{equation}
Aggregate capacity is measured as the two-way within-demeaned median of
$\hat{\lambda}_{it}$ across binding intermediaries, after winsorization at the first
and ninety-ninth percentiles. The shadow value $\hat{\mu}_{it}$ is estimated using an
IV-2SLS procedure that separately identifies loan demand and supply through
predetermined institutional exposures interacted with aggregate funding shocks.

\paragraph{Stage 2: Estimation of dynamic parameters.}
The persistence and capitalization parameters are estimated from
\begin{equation}
  \hat{\lambda}_{i,t+1}
  = \rho\hat{\lambda}_{it} + \psi k_{it} + \alpha_i + \tau_t + \varepsilon_{it},
  \label{eq:panel_ar}
\end{equation}
using within-estimator OLS with institution-clustered standard errors. Aggregate crisis
effects $\hat{\alpha}_{\lambda}$ and policy elasticities $\hat{\gamma}_{\Xi}$ are
recovered from time-series regressions of the cross-sectional median capacity measure
on crisis and policy indicators.

\paragraph{Stage 3: Dynamic responses and treatment effects.}
Dynamic responses are estimated using local projections:
\begin{equation}
  \hat{\lambda}_{i,t+h} - \hat{\lambda}_{i,t-1}
  = \alpha_i^{(h)} + \beta_h C_t + \Gamma_h^{\top} X_{it} + u_{i,t+h},
  \label{eq:lp}
\end{equation}
where $X_{it}$ includes balance-sheet and profitability controls. The sequence
$\{\hat{\beta}_h\}$ provides a nonparametric estimate of the impulse-response function
and is compared with the model-implied response $\hat{\alpha}_{\lambda}\hat{\rho}^{h}$.

The dynamic intention-to-affect treatment effect (DIATE) is
\begin{equation}
  \widehat{\mathrm{DIATE}}(h)
  = \overline{
      \log L^{*}\!\left(\hat{\lambda}^{(1)}_{t+h}\right)
      - \log L^{*}\!\left(\hat{\lambda}^{(0)}_{t+h}\right)
    },
  \label{eq:diate}
\end{equation}
where $\hat{\lambda}^{(0)}$ denotes the no-crisis counterfactual path. Estimation is
restricted to observations with binding constraints, and inference is based on a
500-replication institution-level cluster bootstrap.

\section{Data}
\label{sec:data}

\subsection{Sample Construction}

The analysis combines supervisory balance-sheet data for commercial banks and mutual financial intermediaries in the United States and Brazil. The U.S. sample spans 2002Q4--2025Q2 and is constructed from FDIC Call Reports, Uniform Bank Performance Reports (UBPR), and NCUA regulatory filings. The Brazilian sample spans 2003Q1--2024Q4 and is obtained from the Banco Central do Brasil's IF.Data system under the COSIF reporting framework.

The empirical design exploits the institutional structure of each financial system. In the United States, the ten largest commercial banks by total assets ("Big 10") serve as the dominant providers of wholesale funding and liquidity services, while federally insured credit unions constitute the mutual sector. In Brazil, the nine largest commercial banks ("Big 9") are matched with 137 credit cooperatives that rely on similar wholesale funding relationships. This structure provides cross-sectional variation in intermediary funding capacity that is central to identification.

For each institution, we collect information on lending activity, balance-sheet composition, funding structure, capitalization, profitability, and asset quality. Specifically, we observe total loans ($L_{it}$), deposits ($D_{it}$), wholesale borrowings ($BL_{it}$), equity capital ($K_{it}$), total assets, interest income and expense, and nonperforming loans. All nominal variables are converted to real terms using the GDP deflator for the United States and the IPCA index for Brazil.

\subsection{Variable Construction}

The empirical counterparts of the model variables are constructed from standard regulatory reporting categories. Total loans ($L_{it}$), deposits ($D_{it}$), wholesale borrowings ($BL_{it}$), and equity capital ($K_{it}$) are obtained directly from supervisory balance-sheet schedules. The funding-cost measure, $\omega_{it}$, is proxied by interest expense divided by average interest-bearing liabilities, while the loan rate $r^L_{it}$ is measured as interest income divided by average loans.

To reduce the influence of reporting anomalies and extreme observations, all continuous variables are winsorized at the first and ninety-ninth percentiles within country-quarter cells. Institution and quarter fixed effects are incorporated throughout the analysis using two-way within transformations.

\subsection{Summary Statistics}

Tables~\ref{tab:summary_us} and \ref{tab:summary_brazil} report descriptive statistics for the major commercial banks in each country. Several patterns emerge.

First, U.S. institutions exhibit substantially higher deposit funding ratios and lower net interest margins than their Brazilian counterparts, reflecting differences in financial structure and monetary environments. Second, Brazilian banks display greater variation in lending growth and profitability, consistent with the more cyclical nature of credit provision in emerging-market banking systems. Third, measures of balance-sheet composition and asset quality vary considerably across institutions in both countries, providing meaningful cross-sectional variation for identifying intermediary-specific lending capacity.

Most importantly for the model, both samples exhibit substantial heterogeneity in funding structures, capitalization, and wholesale borrowing intensity. This variation generates the cross-sectional dispersion in effective lending capacity required to recover the latent parameter $\lambda_{it}$ and to estimate its dynamic evolution over time.

\begin{table}[htbp]
\centering
\caption{Summary Statistics: Top 10 U.S.\ Commercial Banks, 2002Q4--2025Q2}
\label{tab:summary_us}
\begin{threeparttable}
\smallskip
\textsc{Panel A: Time-Series Means}
\smallskip
 
\begin{tabular}{%
  l                    
  S[table-format= 3.2] 
  S[table-format= 2.2] 
  S[table-format= 1.2] 
  S[table-format= 1.2] 
  S[table-format= 1.3] 
  S[table-format= 2.3] 
}
\toprule
 & {Loan} & {Loan-to-} & {RE Loan} & {Deposit} & {Net Int.} & {NPL} \\
 & {Growth} & {Asset (\%)} & {Share} & {Ratio} & {Margin} & {Ratio} \\
\midrule
Bank of America & 169.99 & 70.51 &  1.00 & 0.55 & 0.026 & -0.140 \\
Capital One     &  23.49 &  1.79 &  0.37 & 0.90 & 0.045 &  3.450 \\
Citibank        &   4.46 &  2.85 &  0.29 & 0.78 & 0.032 &  0.560 \\
Goldman Sachs   &  10.43 &  1.08 &  0.19 & 0.61 & 0.009 &  0.100 \\
JPMorgan Chase  &  10.91 & 34.51 &  0.44 & 0.71 & 0.022 &  0.400 \\
PNC Bank        &  11.54 &  3.23 &  0.44 & 0.83 & 0.032 & -3.550 \\
State Street    &  11.37 & -3.41 &  0.03 & 0.80 & 0.015 & -0.003 \\
Truist Bank     &  11.52 &  1.03 &  0.60 & 0.83 & 0.034 &  2.020 \\
US Bank         &   5.50 &  0.89 &  0.47 & 0.81 & 0.034 &  0.850 \\
Wells Fargo     &  12.06 &  0.06 &  0.56 & 0.86 & 0.036 & -0.750 \\
\bottomrule
\end{tabular}
 
\bigskip
 
\textsc{Panel B: Time-Series Standard Deviations}
\smallskip
 
\begin{tabular}{%
  l
  S[table-format= 3.2]
  S[table-format= 2.2]
  S[table-format= 1.3]
  S[table-format= 1.2]
  S[table-format= 1.3]
  S[table-format= 2.2]
}
\toprule
 & {Loan} & {Loan-to-} & {RE Loan} & {Deposit} & {Net Int.} & {NPL} \\
 & {Growth} & {Asset (\%)} & {Share} & {Ratio} & {Margin} & {Ratio} \\
\midrule
Bank of America & 601.17 & 20.70 & 0.009 & 0.43 & 0.011 &  0.93 \\
Capital One     &  62.73 &  6.60 & 0.163 & 0.04 & 0.010 & 28.73 \\
Citibank        &  13.69 & 19.01 & 0.096 & 0.06 & 0.005 &  2.95 \\
Goldman Sachs   &  47.50 & 14.41 & 0.052 & 0.23 & 0.006 &  0.40 \\
JPMorgan Chase  &  21.63 &  3.24 & 0.069 & 0.10 & 0.004 &  2.71 \\
PNC Bank        &  23.64 & 19.48 & 0.070 & 0.05 & 0.006 & 33.47 \\
State Street    &  21.65 & 44.21 & 0.032 & 0.08 & 0.004 &  1.17 \\
Truist Bank     &  24.05 &  1.64 & 0.136 & 0.07 & 0.003 & 12.05 \\
US Bank         &   4.21 &  0.93 & 0.033 & 0.07 & 0.005 &  3.34 \\
Wells Fargo     &  24.84 &  3.68 & 0.083 & 0.06 & 0.007 &  5.82 \\
\bottomrule
\end{tabular}
 
\begin{tablenotes}[flushleft]
\footnotesize
\item \textit{Notes.} Loan Growth is the quarterly percentage change in total loans.
  Loan-to-Asset is total loans divided by total assets (expressed as a percentage).
  RE Loan Share is real-estate loans divided by total loans.
  Deposit Ratio is total deposits divided by total liabilities.
  Net Int.\ Margin is net interest income divided by earning assets.
  NPL Ratio is nonperforming loans divided by total loans.
  Source: FDIC Call Reports.
\end{tablenotes}
\end{threeparttable}
\end{table}
 
\begin{table}[htbp]
\centering
\caption{Summary Statistics: Top 9 Brazilian Commercial Banks, 2003Q1--2024Q4}
\label{tab:summary_brazil}
\begin{threeparttable}
\smallskip
\textsc{Panel A: Time-Series Means}
\smallskip
 
\begin{tabular}{%
  l
  S[table-format= 1.3]
  S[table-format= 1.3]
  S[table-format= 1.3]
  S[table-format= 1.3]
  S[table-format= 1.3]
  S[table-format=-1.3]
}
\toprule
 & {Loan} & {Loan-to-} & {RE Loan} & {Deposit} & {Net Int.} & {NPL} \\
 & {Growth} & {Asset} & {Share} & {Ratio} & {Margin} & {Ratio} \\
\midrule
Banco Coop.\ Sicredi   & 0.049 & 0.313 & 0.247 & 0.305 & 0.177 & -0.001 \\
Banco do Brasil        & 0.031 & 0.394 & 0.188 & 0.349 & 0.193 & -0.058 \\
Bradesco               & 0.027 & 0.349 & 0.140 & 0.264 & 0.268 & -0.082 \\
BTG Pactual            & 0.092 & 0.138 & 0.373 & 0.129 & 0.372 & -0.036 \\
Caixa Econ.\ Federal   & 0.048 & 0.403 & 0.004 & 0.000 & 0.666 & -0.073 \\
Citibank (BR)          & 0.010 & 0.176 & 0.108 & 0.168 & 0.309 & -0.064 \\
Ita\'{u}               & 0.034 & 0.314 & 0.126 & 0.238 & 0.240 & -0.081 \\
Safra                  & 0.025 & 0.345 & 0.103 & 0.129 & 0.458 & -0.038 \\
Santander              & 0.027 & 0.353 & 0.148 & 0.213 & 0.332 & -0.069 \\
\bottomrule
\end{tabular}
 
\bigskip
 
\textsc{Panel B: Time-Series Standard Deviations}
\smallskip
 
\begin{tabular}{%
  l
  S[table-format=1.3]
  S[table-format=1.3]
  S[table-format=1.3]
  S[table-format=1.3]
  S[table-format=1.3]
  S[table-format=1.3]
}
\toprule
 & {Loan} & {Loan-to-} & {RE Loan} & {Deposit} & {Net Int.} & {NPL} \\
 & {Growth} & {Asset} & {Share} & {Ratio} & {Margin} & {Ratio} \\
\midrule
Banco Coop.\ Sicredi   & 0.121 & 0.111 & 0.550 & 0.182 & 0.156 & 0.001 \\
Banco do Brasil        & 0.086 & 0.046 & 0.345 & 0.177 & 0.186 & 0.010 \\
Bradesco               & 0.085 & 0.032 & 0.223 & 0.153 & 0.177 & 0.013 \\
BTG Pactual            & 0.358 & 0.061 & 0.474 & 0.106 & 0.154 & 0.016 \\
Caixa Econ.\ Federal   & 0.095 & 0.154 & 0.033 & 0.000 & 0.026 & 0.029 \\
Citibank (BR)          & 0.103 & 0.066 & 0.203 & 0.099 & 0.118 & 0.049 \\
Ita\'{u}               & 0.086 & 0.025 & 0.203 & 0.127 & 0.186 & 0.016 \\
Safra                  & 0.107 & 0.047 & 0.181 & 0.090 & 0.153 & 0.010 \\
Santander              & 0.155 & 0.032 & 0.126 & 0.135 & 0.171 & 0.009 \\
\bottomrule
\end{tabular}
 
\begin{tablenotes}[flushleft]
\footnotesize
\item \textit{Notes.} Variable definitions mirror Table~\ref{tab:summary_us}.
  All variables are expressed in proportional (not percentage) units following
  the reporting conventions of the COSIF framework.
  Source: Banco Central do Brasil IF.Data.
\end{tablenotes}
\end{threeparttable}
\end{table}
\section{Results}
\label{sec:results}

The results section traces four steps in the narrative. We begin by documenting the
recovered capacity multiplier and how it differs across countries and crises. We then
show how capacity responds dynamically to each shock, using both the model's analytical
IRF and the nonparametric LP-IRF as a check. The DIATE profiles translate capacity
movements into lending effects, episode by episode. Finally, policy counterfactuals ask
how much of the documented cross-country divergence could be closed by backstop
interventions or by institutional deepening.

\subsection{Lending Capacity Across Countries and Crisis Episodes}

We begin by estimating the lending-capacity multiplier, $\hat{\lambda}_{it}$, for institutions operating at the funding constraint. Table~\ref{tab:lambda} reports cross-sectional median estimates by country and crisis episode.

Two findings emerge. First, lending capacity is systematically higher in the United States than in Brazil. Across episodes, U.S. estimates range from 0.45 to 0.63, whereas Brazilian estimates range from 0.10 to 0.18. The implied difference is economically large: the effective lending capacity generated by a unit of wholesale funding is approximately three to six times greater in the United States. Second, the cross-country ranking is remarkably stable across crisis environments. Although the Global Financial Crisis, the post-QE normalization episode, the Brazilian liquidity crisis, and the COVID-19 shock differ substantially in origin and transmission, the relative ordering of lending capacity remains unchanged.

The estimates also reveal a gradual decline in capacity over time within both countries. In the United States, the median multiplier falls from 0.63 during the Global Financial Crisis to 0.45 during the COVID-19 episode. Brazil exhibits a similar pattern, with the multiplier declining from 0.18 to 0.10 over the same period. These results suggest a secular reduction in effective wholesale funding capacity, consistent with tighter balance-sheet constraints and diminished intermediary funding flexibility.

Viewed through the lens of the model, the estimated multipliers imply substantial differences in steady-state lending capacity. Because lending is constrained by $D_{it}+\lambda_{it}BL_{it}$, higher values of $\lambda$ mechanically expand the amount of credit that can be supported by a given funding base. The evidence therefore points to a structurally deeper intermediary sector in the United States and a substantially thinner capacity buffer in Brazil.

\begin{table}[htbp]
\centering
\begin{threeparttable}
\resizebox{\textwidth}{!}{
\begin{tabular}{lcccccc}
\toprule\toprule
 & \multicolumn{3}{c}{United States} & \multicolumn{3}{c}{Brazil} \\
\cmidrule(lr){2-4}\cmidrule(lr){5-7}
 & GFC & QE III & COVID-19 & GFC & Liquidity & COVID-19 \\
\midrule
$\hat\lambda$ (Capacity Multiplier) & 0.632$^{***}$ & 0.498$^{***}$ & 0.450$^{***}$ & 0.175$^{***}$ & 0.174$^{**}$ & 0.097$^{*}$ \\
Standard Error & (0.021) & (0.026) & (0.035) & (0.046) & (0.066) & (0.054) \\
Observations & 424,494 & 424,494 & 424,494 & 6,907 & 3,428 & 1,868 \\
\bottomrule\bottomrule
\end{tabular}}
\begin{tablenotes}[flushleft]
\small
\item \textit{Notes.} Estimates are median values of $\hat\lambda_{it}=(L_{it}-D_{it})/BL_{it}$ restricted to binding
institutions ($\hat\mu_{it}>0$), two-way within-demeaned (institution $\times$ quarter). Sample is winsorized
at the 1\textsuperscript{st} and 99\textsuperscript{th} percentiles. Standard errors clustered by institution.
$^{***}p<0.01$, $^{**}p<0.05$, $^{*}p<0.10$.
\end{tablenotes}
\end{threeparttable}
\caption{Static Loan-Capacity Multiplier ($\hat\lambda$) by Country and Crisis Episode.}
\label{tab:lambda}
\end{table}

Table~\ref{tab:lambda} establishes the first empirical prediction of the model: economies characterized by higher lending capacity enter crisis episodes with fewer institutions operating at the binding margin and, consequently, greater resilience to adverse funding shocks.

\subsection{Dynamic Responses of Lending Capacity}

\begin{figure}[htbp]
\centering
\includegraphics[width=0.5\textwidth]{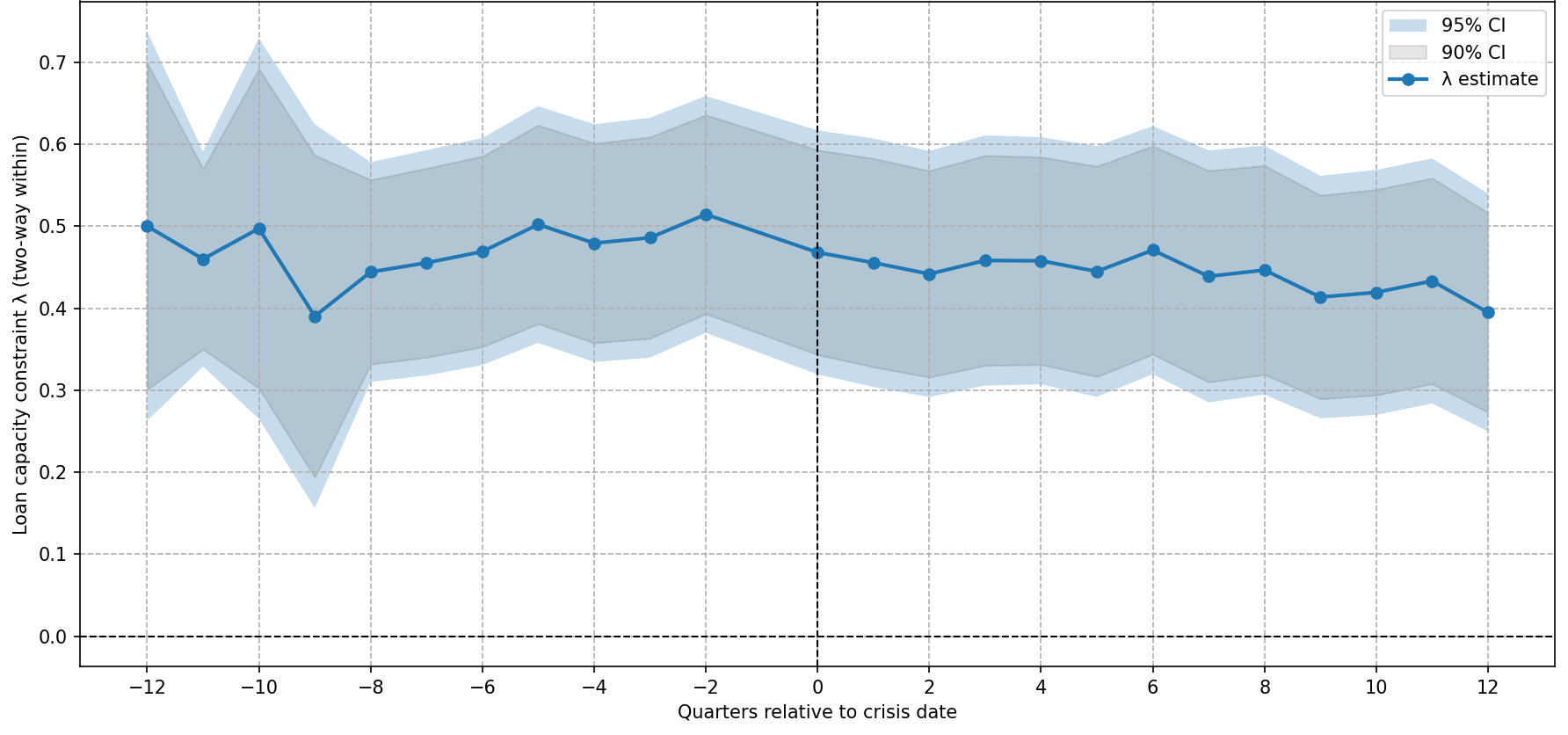}\hfill
\includegraphics[width=0.5\textwidth]{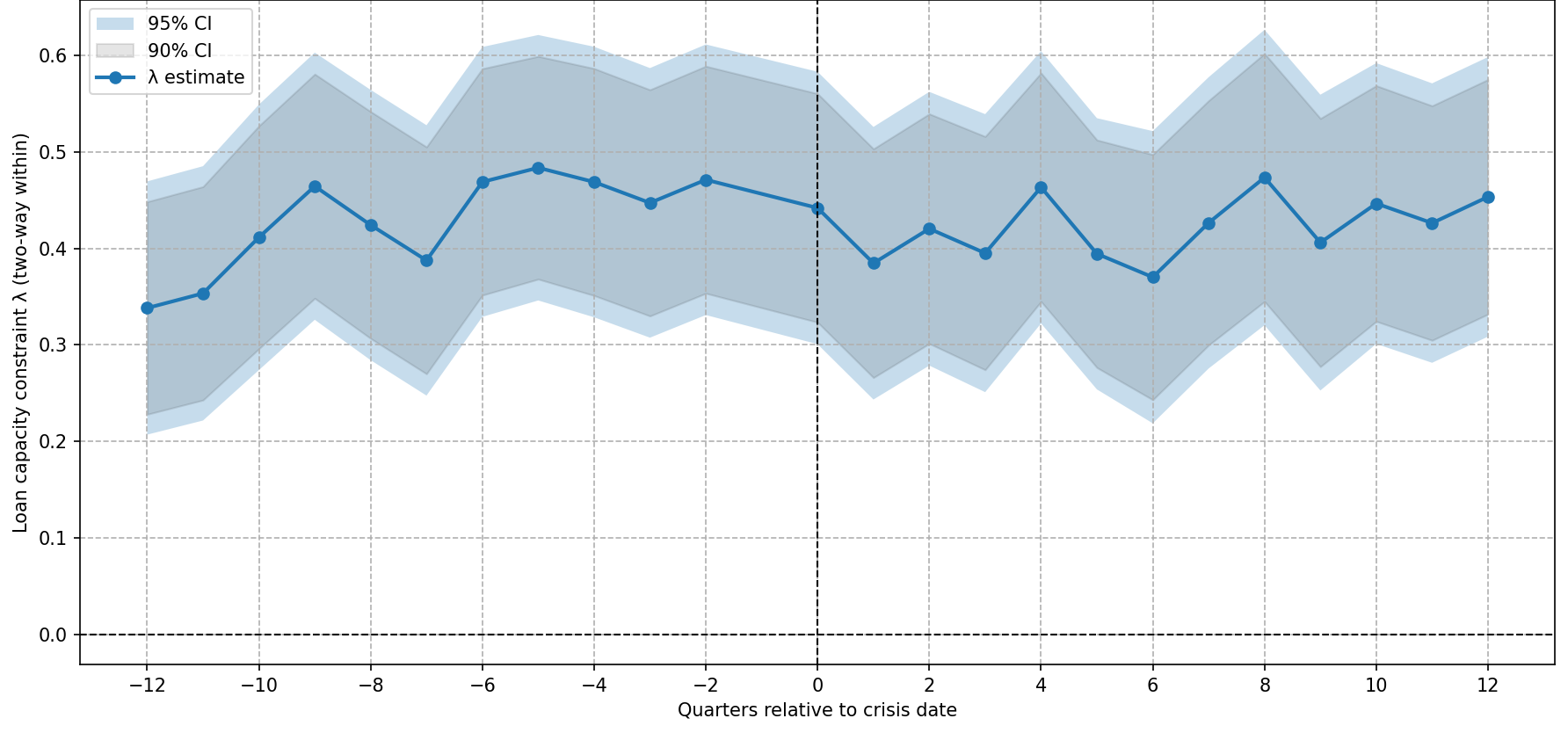}

\parbox{0.48\textwidth}{\small\centering\textit{GFC (2008Q4)}}%
\hfill
\parbox{0.48\textwidth}{\small\centering\textit{End of QE~III (2014Q3)}}

\vspace{1em}

\includegraphics[width=0.5\textwidth]{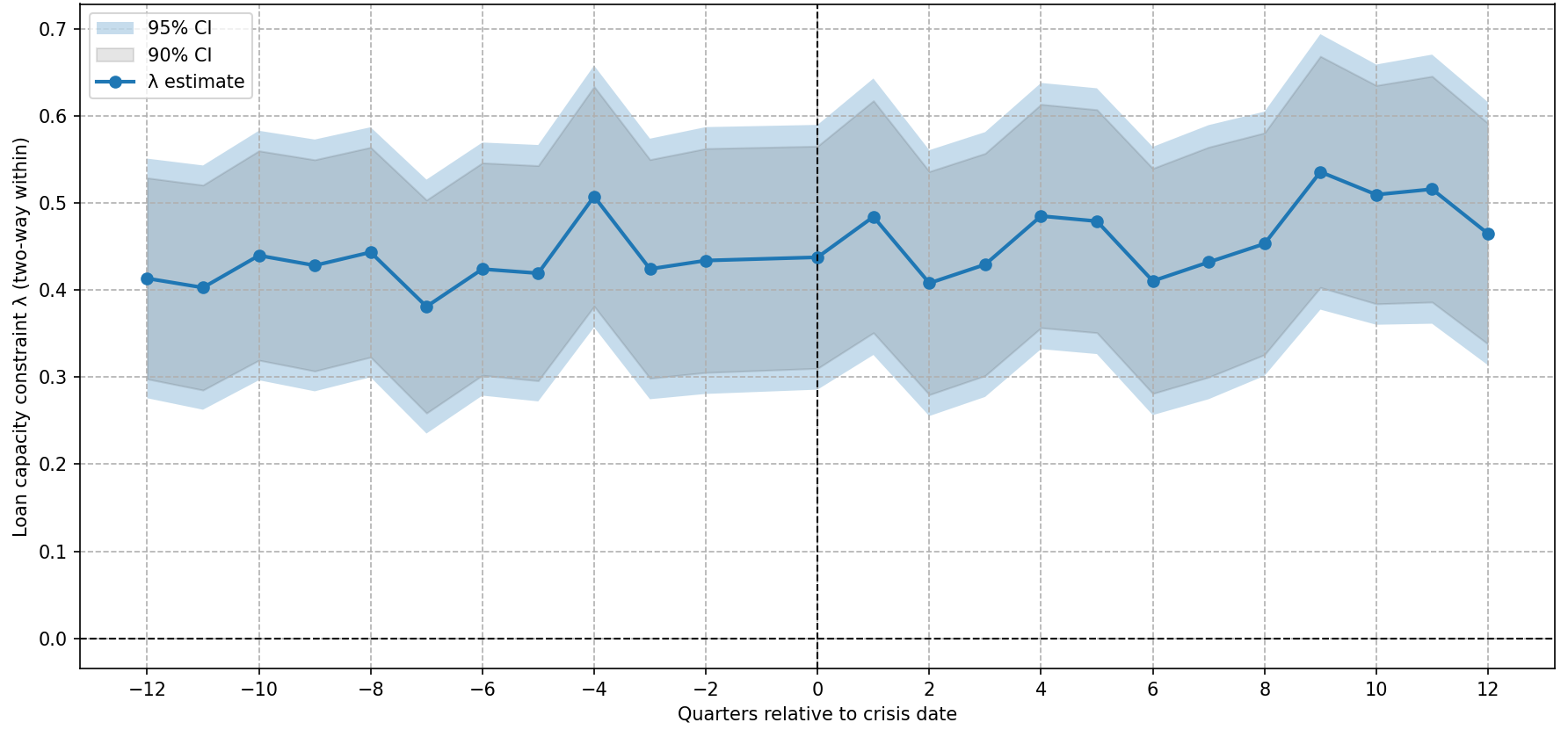}

\parbox{0.48\textwidth}{\small\centering\textit{COVID-19 (2020Q1)}}

\caption{U.S.\ Event Study of Loan-Capacity Multiplier $\hat{\lambda}_t$. Each panel
plots $\hat{\beta}_k$ from the event-study regression with $k=-12,\ldots,12$, with 90\%
(light) and 95\% (dark) confidence bands. Estimates are two-way within-demeaned.
The omitted bin is $k=-1$. Standard errors clustered by institution.}
\label{fig:event_us}
\end{figure}

\begin{figure}[htbp]
\centering
\includegraphics[width=0.5\textwidth]{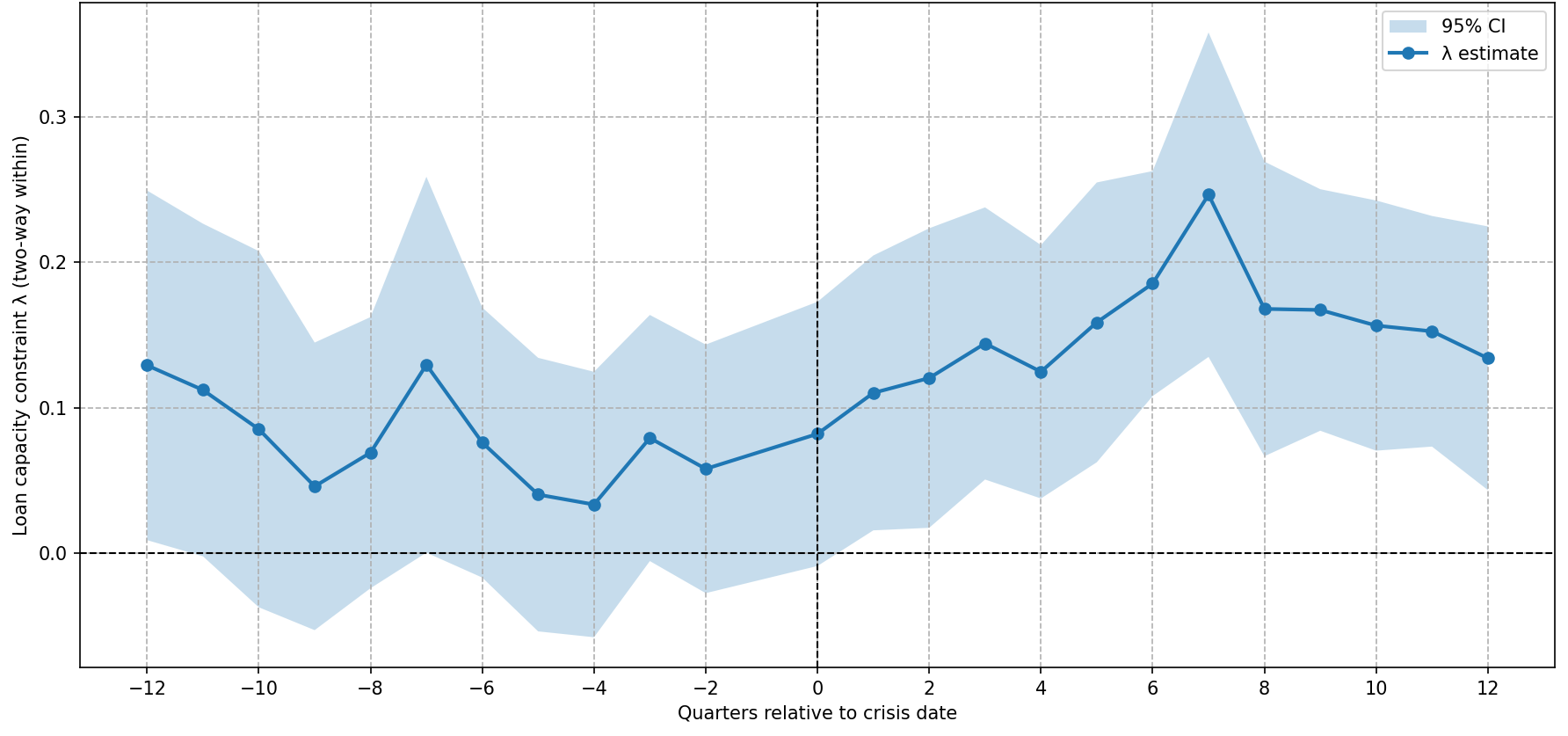}\hfill
\includegraphics[width=0.5\textwidth]{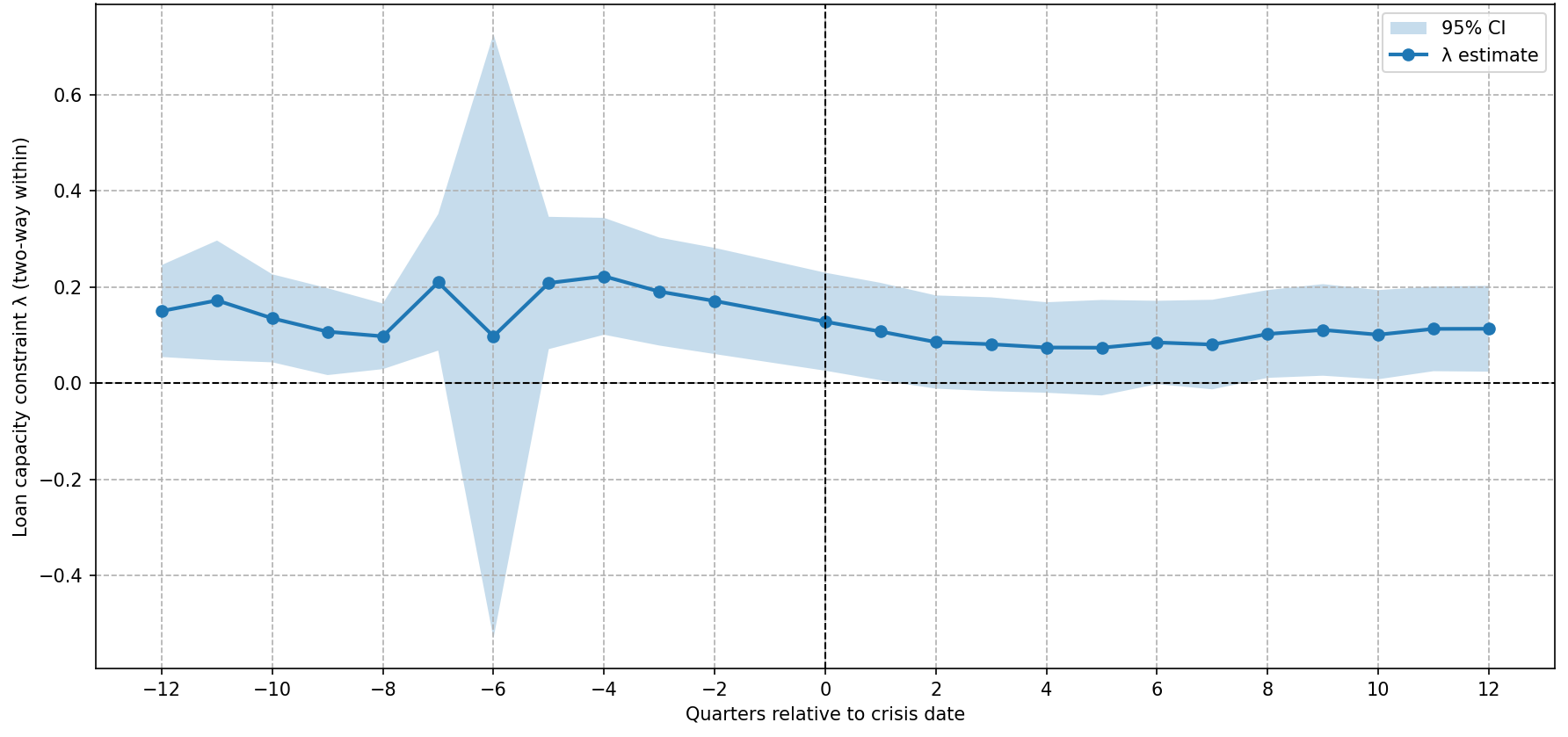}

\parbox{0.48\textwidth}{\small\centering\textit{GFC (2008Q4)}}%
\hfill
\parbox{0.48\textwidth}{\small\centering\textit{Liquidity Crisis (2015Q4)}}

\vspace{1em}

\includegraphics[width=0.5\textwidth]{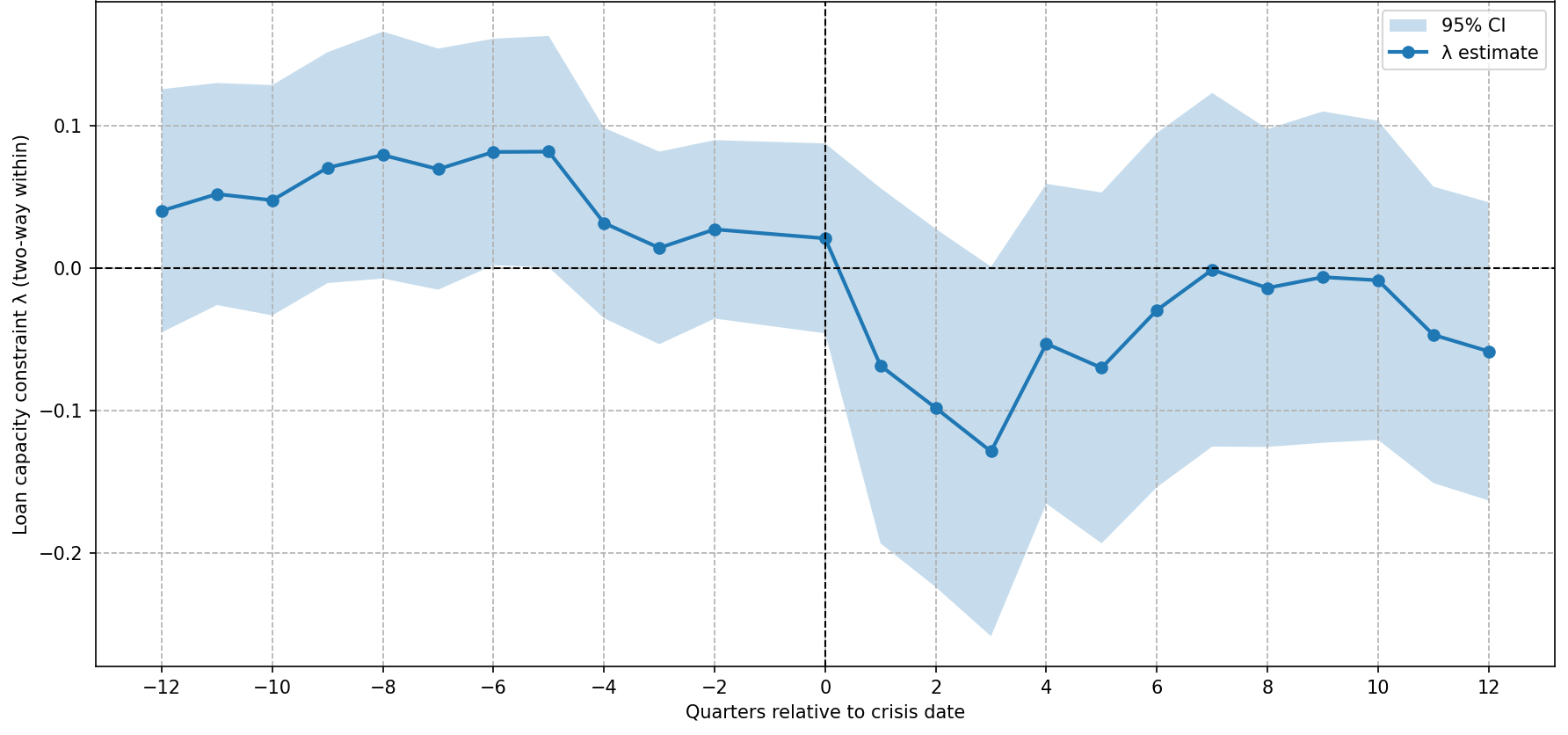}

\parbox{0.48\textwidth}{\small\centering\textit{COVID-19 (2020Q1)}}

\caption{Brazil Event Study of Loan-Capacity Multiplier $\hat{\lambda}_t$.
Notes as in Figure~\ref{fig:event_us}.}
\label{fig:event_brazil}
\end{figure}

\begin{figure}[htbp]
\centering
\includegraphics[width=0.5\textwidth]{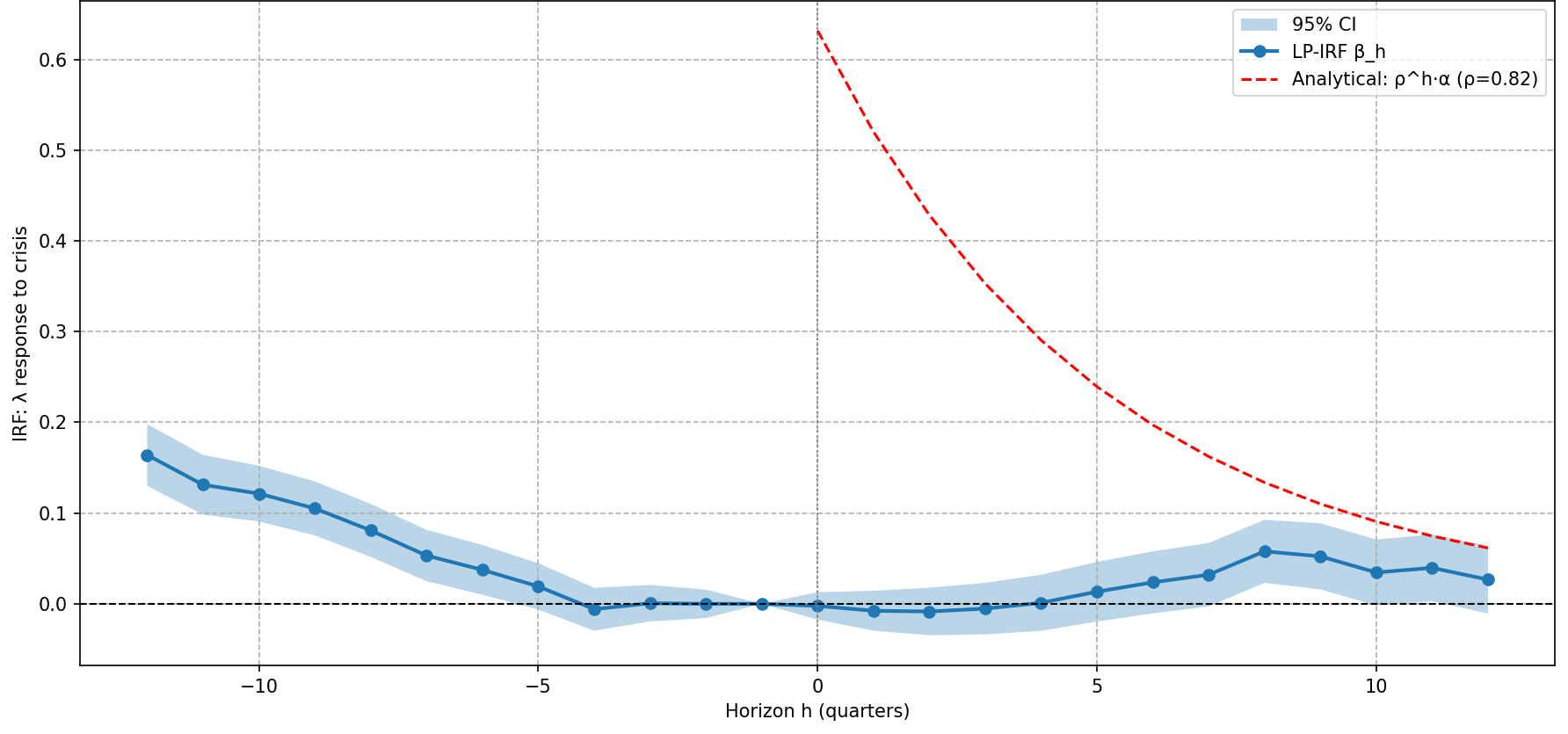}\hfill
\includegraphics[width=0.5\textwidth]{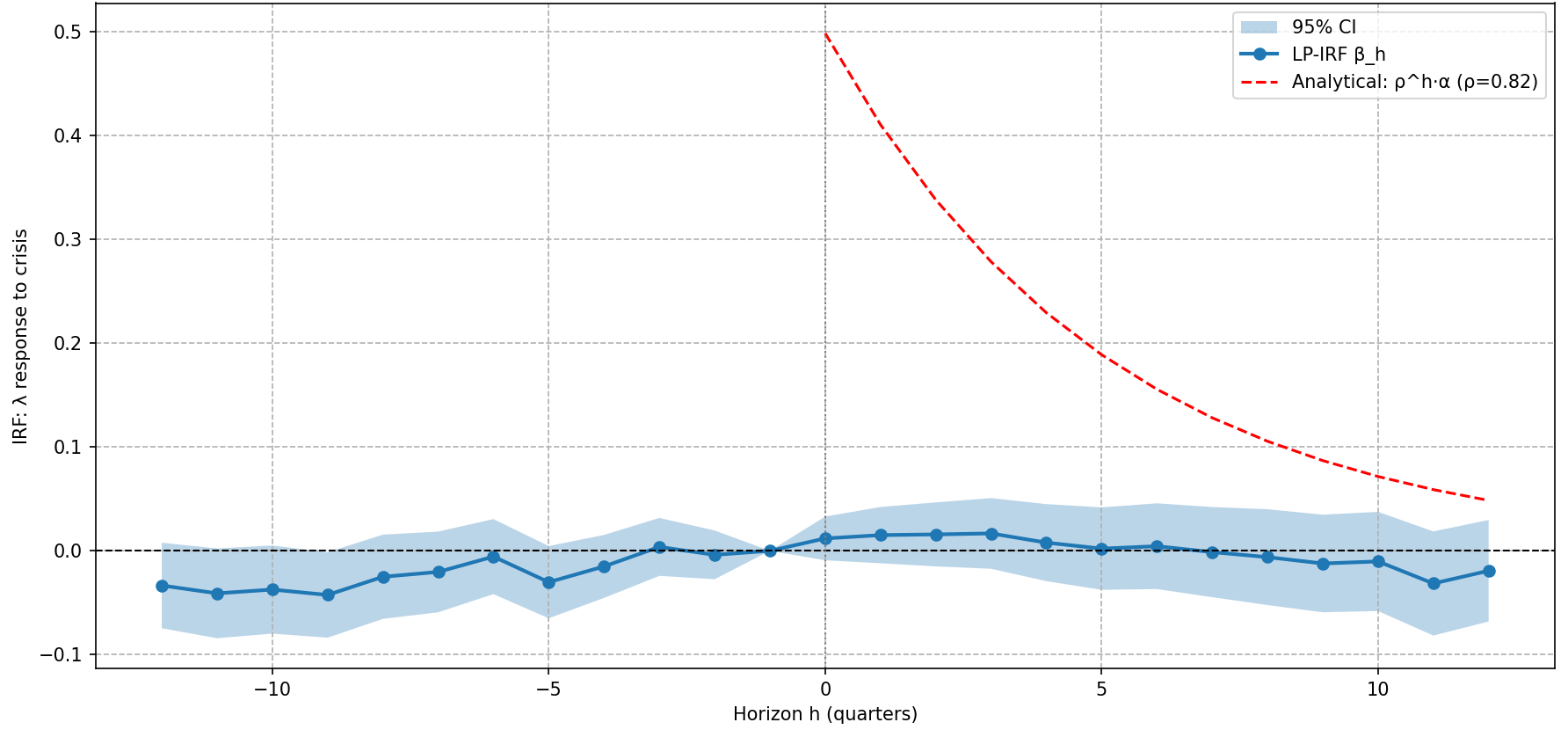}

\parbox{0.48\textwidth}{\small\centering\textit{GFC (2008Q4)}}%
\hfill
\parbox{0.48\textwidth}{\small\centering\textit{End of QE~III (2014Q3)}}

\vspace{1em}

\includegraphics[width=0.5\textwidth]{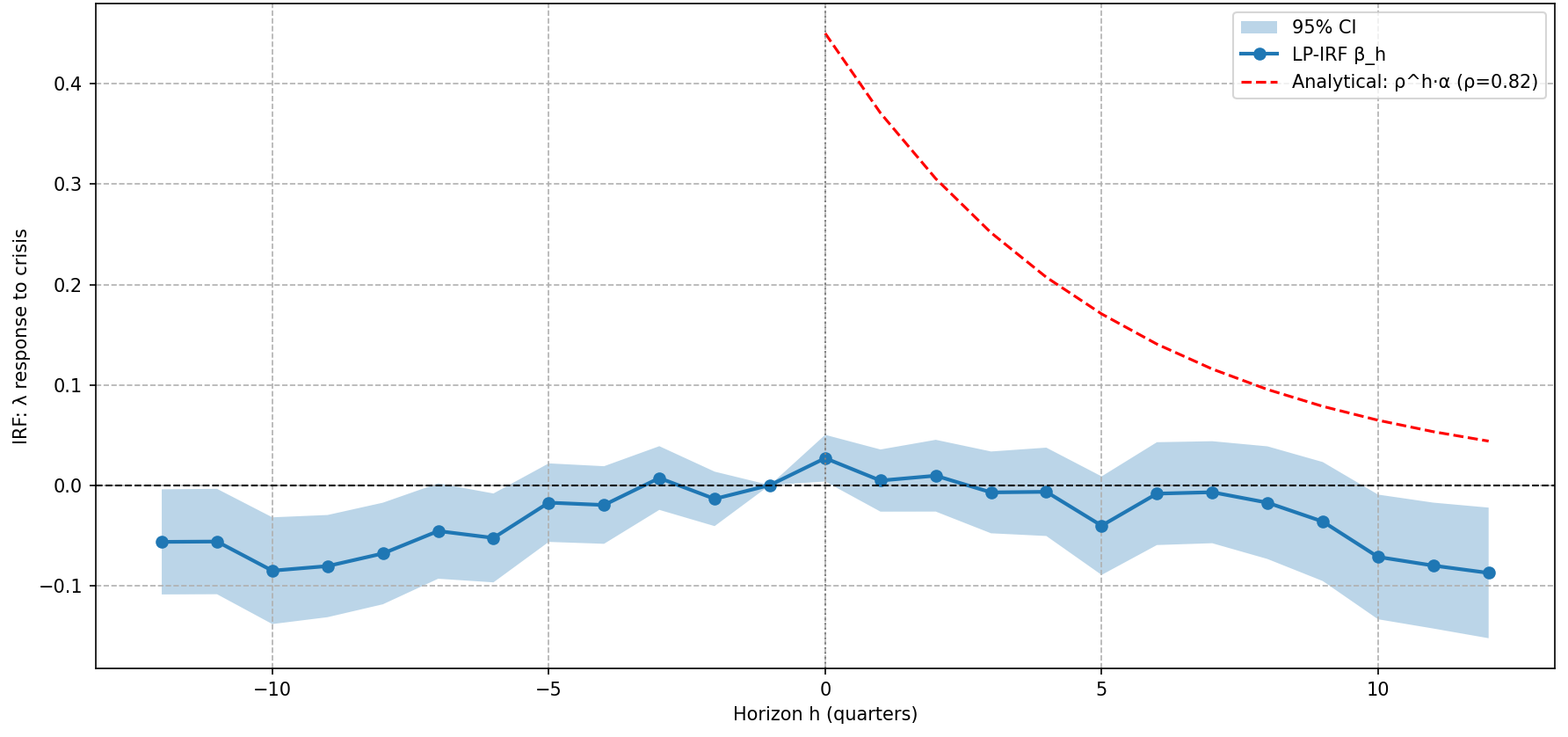}

\parbox{0.48\textwidth}{\small\centering\textit{COVID-19 (2020Q1)}}

\caption{U.S.\ Local Projection IRF for $\hat{\lambda}_{it}$. Solid circles are
$\hat{\beta}_h$ from~\eqref{eq:lp}. Light shaded region is 95\% confidence band;
dark shaded is 90\%. Dashed line is the analytical IRF
$\hat{\alpha}_{\lambda}\hat{\rho}^{h}$. Standard errors clustered by institution.}
\label{fig:lp_us}
\end{figure}

\begin{figure}[htbp]
\centering
\includegraphics[width=0.5\textwidth]{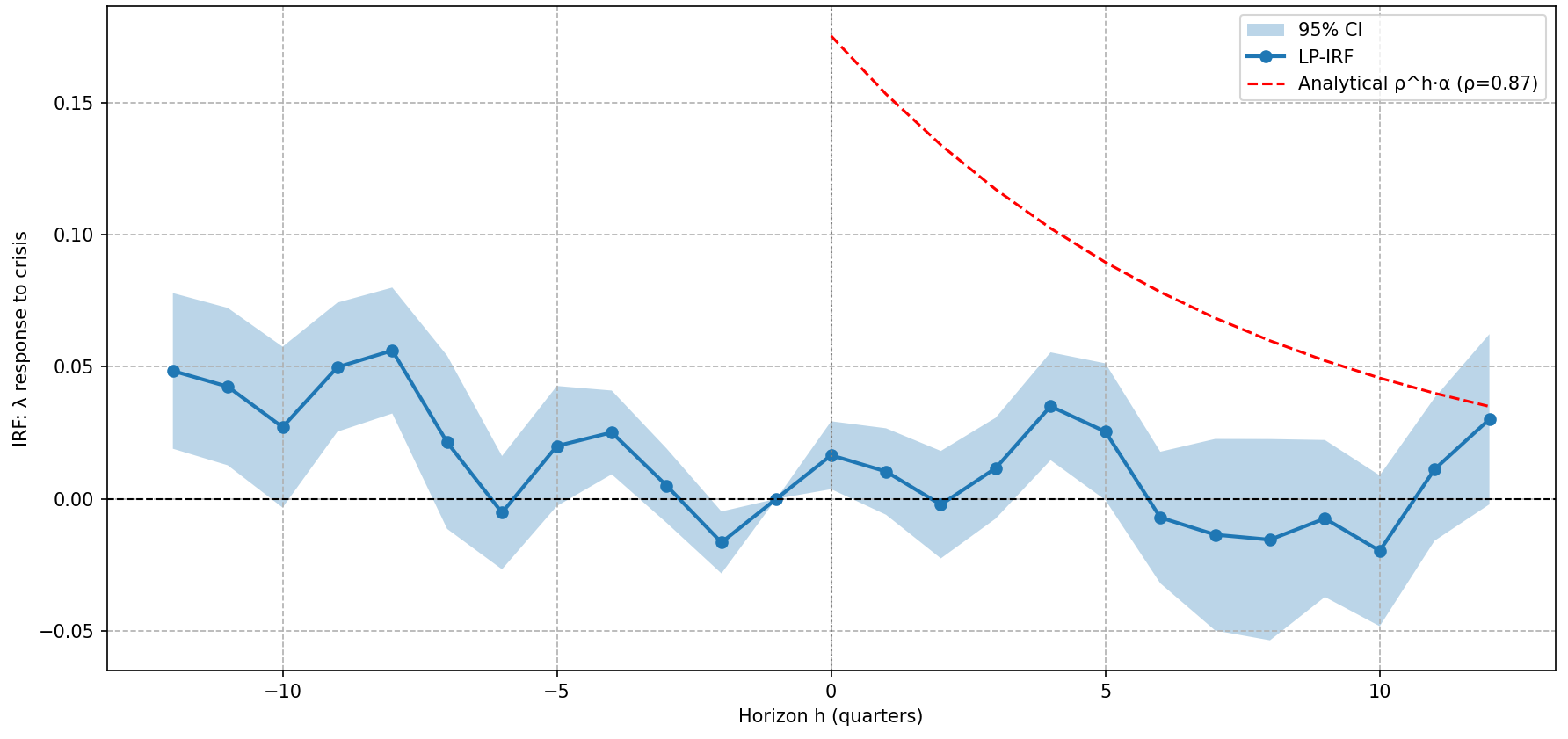}\hfill
\includegraphics[width=0.5\textwidth]{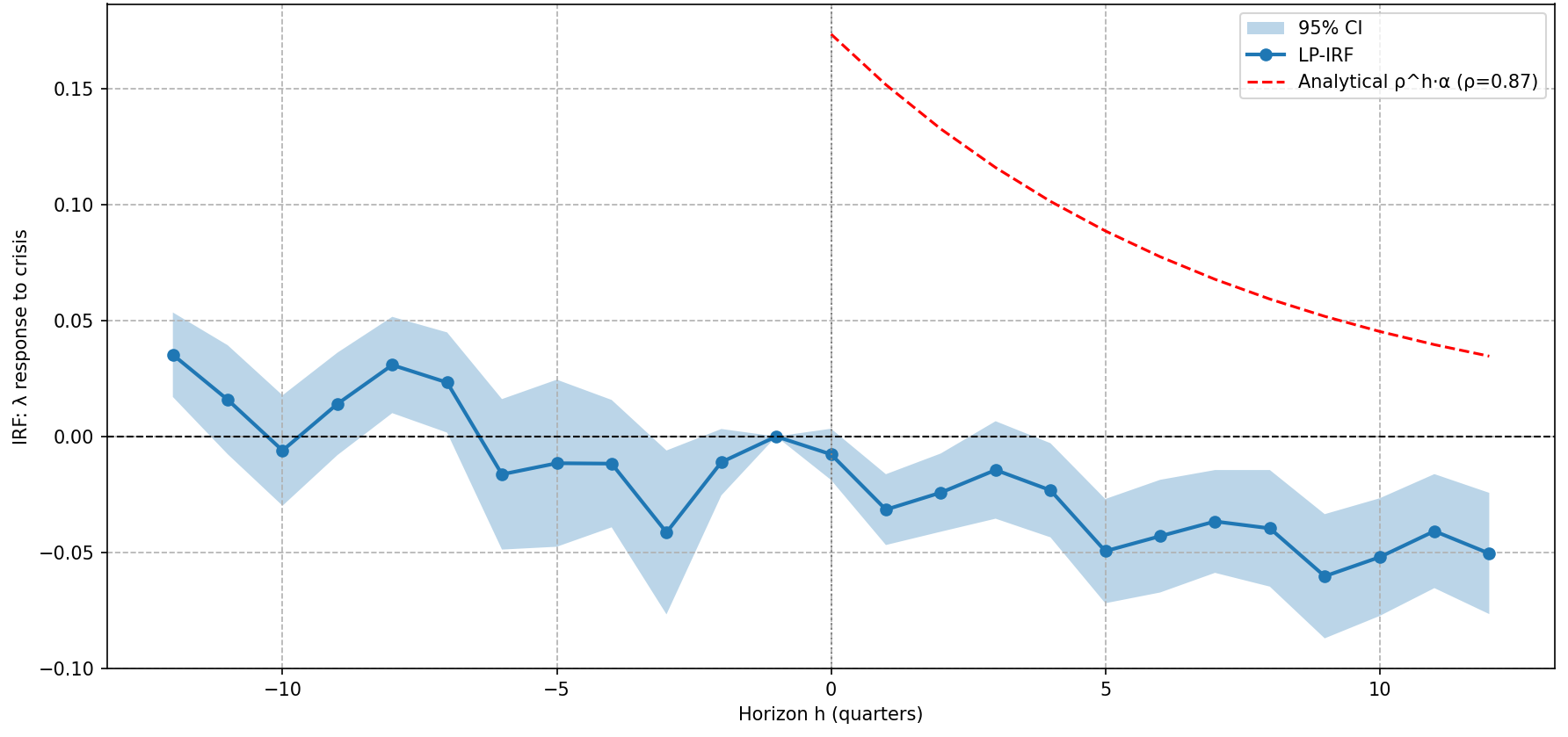}

\parbox{0.48\textwidth}{\small\centering\textit{GFC (2008Q4)}}%
\hfill
\parbox{0.48\textwidth}{\small\centering\textit{Liquidity Crisis (2015Q4)}}

\vspace{1em}

\includegraphics[width=0.5\textwidth]{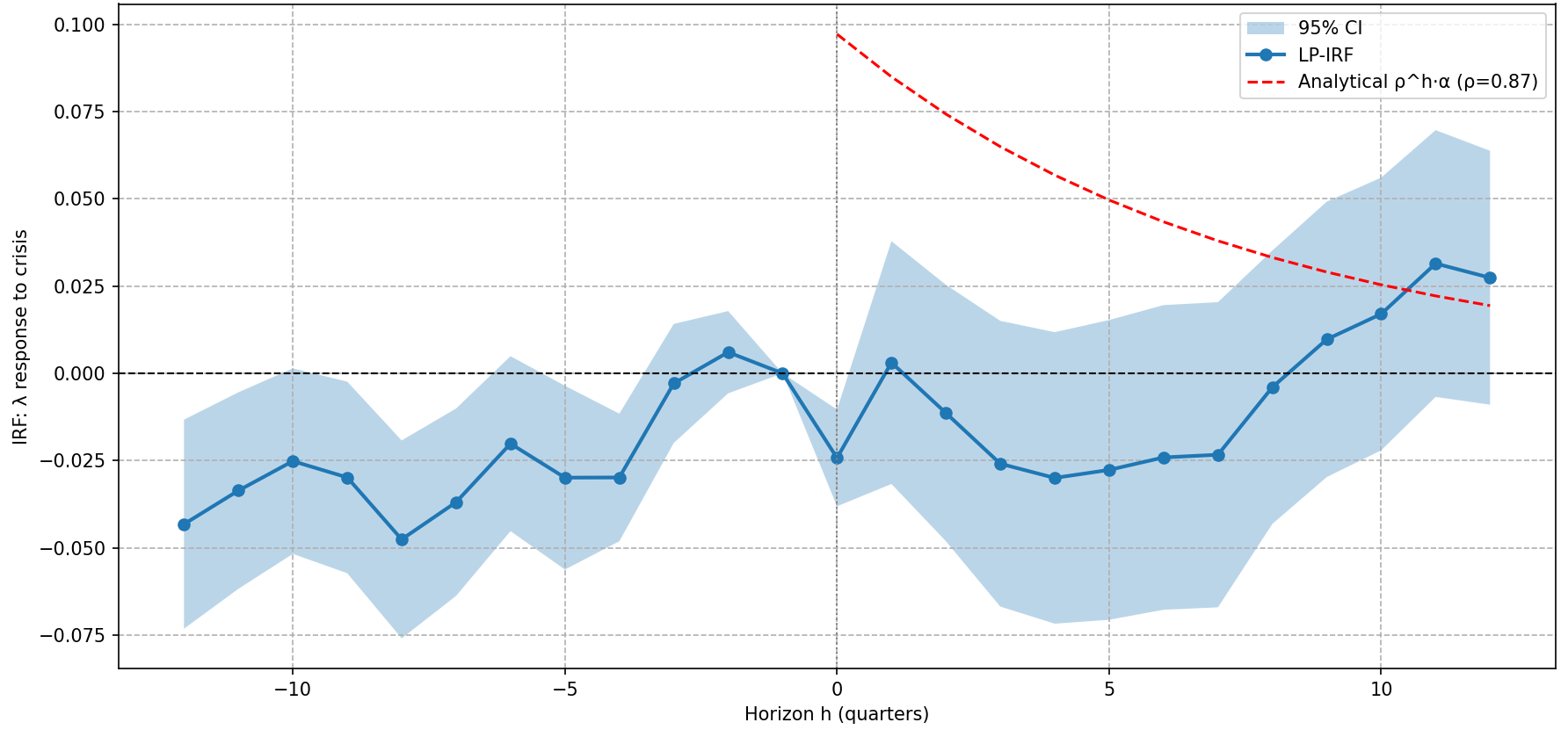}

\parbox{0.48\textwidth}{\small\centering\textit{COVID-19 (2020Q1)}}

\caption{Brazil Local Projection IRF for $\hat{\lambda}_{it}$.
Notes as in Figure~\ref{fig:lp_us}.}
\label{fig:lp_brazil}
\end{figure}

We next examine the dynamic response of lending capacity to crisis shocks. Figures~\ref{fig:event_us}--\ref{fig:event_brazil} present event-study estimates of the aggregate capacity measure, while Figures~\ref{fig:lp_us}--\ref{fig:lp_brazil} report local-projection impulse responses. Throughout, the analytical impulse response implied by the estimated law of motion,
$\hat{\alpha}_\lambda \hat{\rho}^{,h}$, is overlaid as a model-based benchmark.

The event-study evidence reveals substantial differences in post-shock adjustment across countries. In the United States, lending capacity remains comparatively stable around each crisis episode. Although capacity declines modestly at the onset of stress events, the estimated effects are short-lived and recovery occurs rapidly. No episode generates a prolonged deterioration in intermediary funding capacity.

Brazil exhibits markedly different dynamics. Capacity declines are larger, more persistent, and more heterogeneous across episodes. Following the 2015 liquidity crisis and the COVID-19 shock, lending capacity remains significantly below its pre-crisis level throughout most of the post-event horizon. In contrast, the post-2008 episode is characterized by a comparatively rapid recovery, suggesting that policy intervention partially offset the initial contraction in intermediary funding capacity.

The local-projection estimates reinforce these conclusions. For the United States, impulse responses are generally small in magnitude and revert toward baseline within a relatively short horizon. In several episodes, point estimates become positive during the recovery phase, indicating that policy support and balance-sheet repair more than offset the initial disruption. Importantly, pre-event coefficients are statistically indistinguishable from zero, supporting the identifying assumption that crisis timing is orthogonal to pre-existing trends in lending capacity.

For Brazil, the estimated impulse responses are substantially larger and more persistent. The 2015 liquidity crisis generates a prolonged decline in capacity that deepens during the first several quarters following the shock before gradually reverting. The COVID-19 episode produces an immediate contraction that remains statistically significant over much of the forecast horizon. These results indicate that adverse funding shocks propagate through intermediary balance sheets for considerably longer in Brazil than in the United States.

A final result is that the analytical impulse response implied by the model closely tracks the nonparametric local-projection estimates in both countries. The close correspondence between $\hat{\alpha}_\lambda \hat{\rho}^{,h}$ and the estimated LP-IRFs suggests that the parsimonious AR(1) specification for lending capacity provides a reasonable approximation to the observed dynamics. This finding supports the model's interpretation of crisis propagation as operating primarily through persistent shifts in intermediary funding capacity.

\subsection{Dynamic Lending Effects of Capacity Shocks}

\begin{figure}[htbp]
\centering
\includegraphics[width=0.5\textwidth]{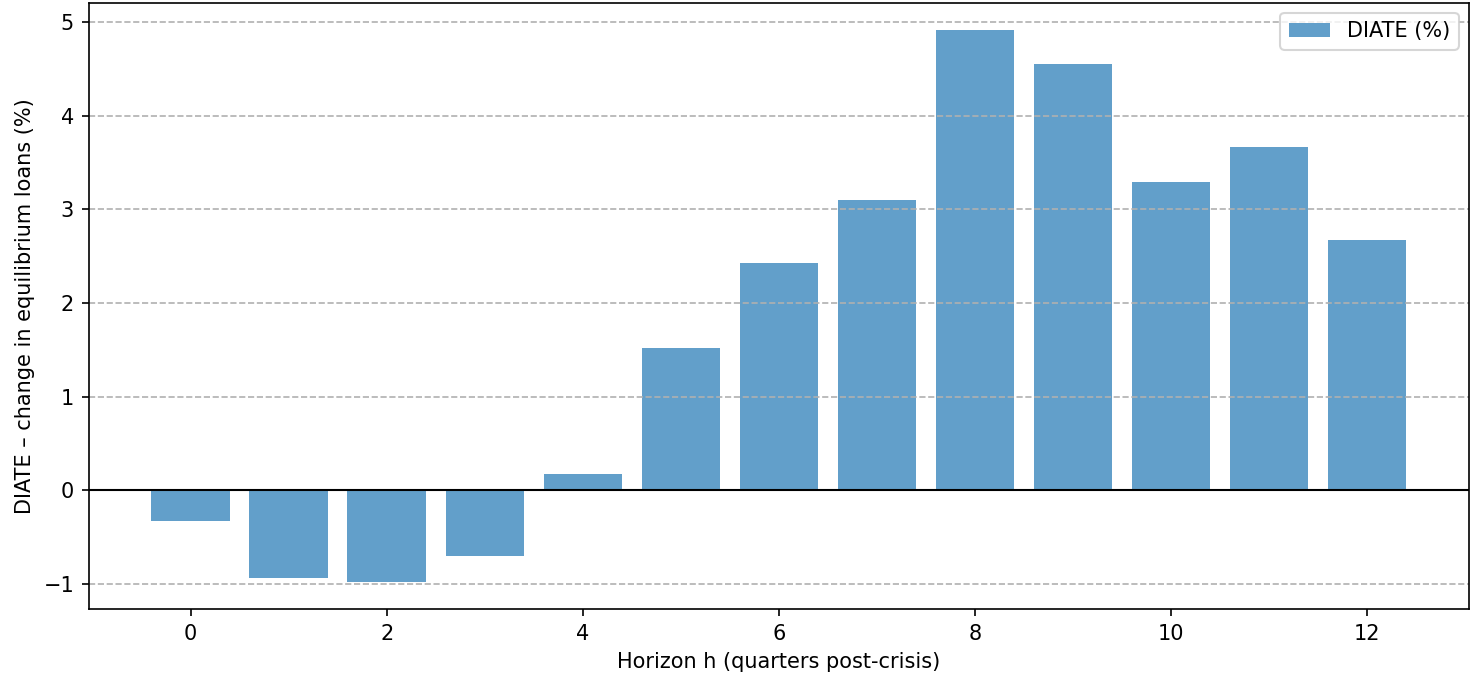}\hfill
\includegraphics[width=0.5\textwidth]{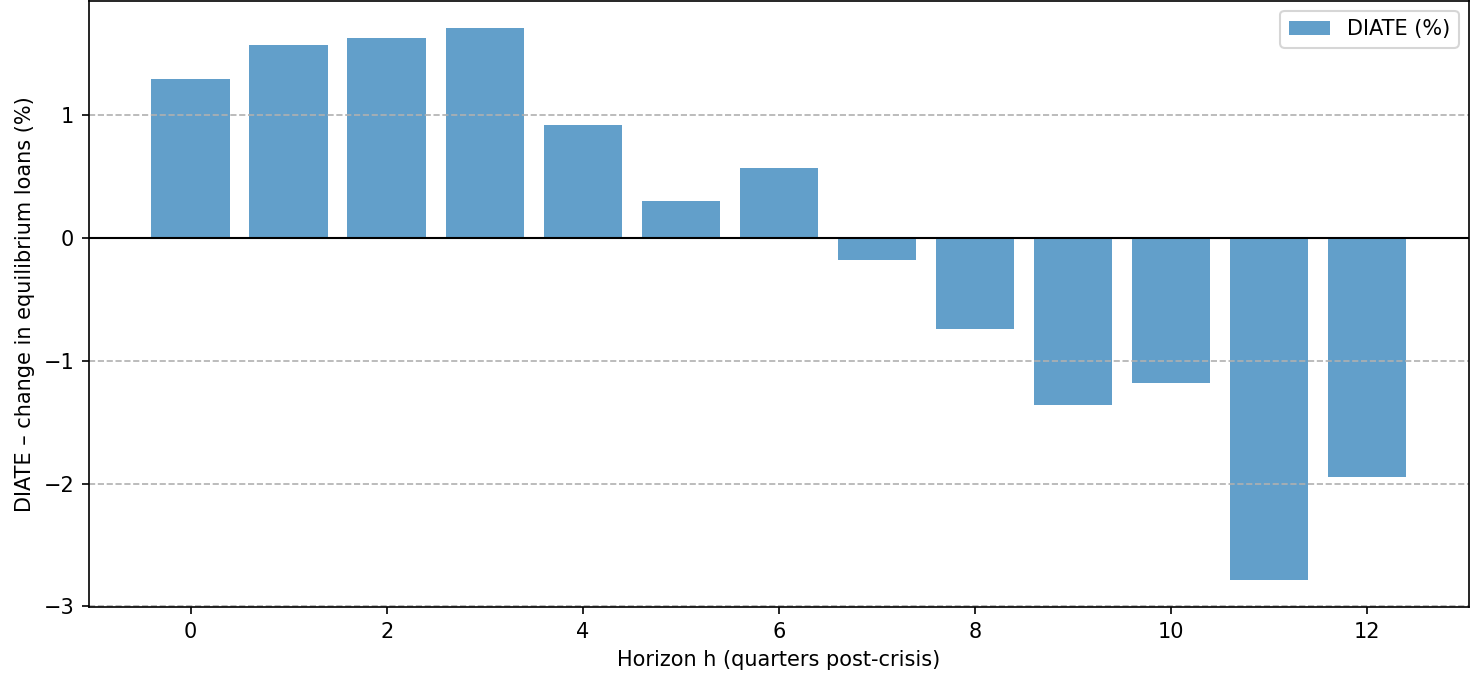}

\parbox{0.48\textwidth}{\small\centering\textit{GFC (2008Q4)}}%
\hfill
\parbox{0.48\textwidth}{\small\centering\textit{End of QE~III (2014Q3)}}

\vspace{1em}

\includegraphics[width=0.5\textwidth]{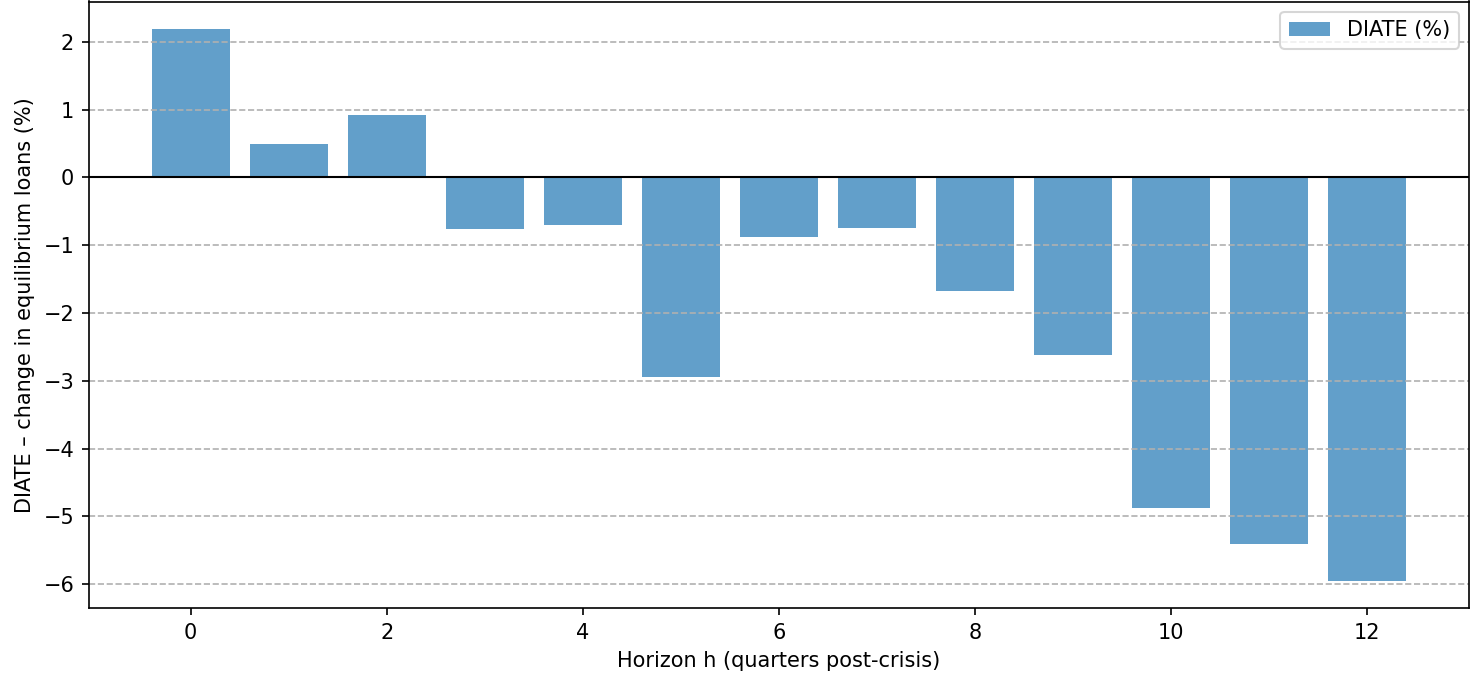}

\parbox{0.48\textwidth}{\small\centering\textit{COVID-19 (2020Q1)}}

\caption{U.S.\ DIATE Profile $\widehat{\mathrm{DIATE}}(h)$ at horizons $h=0,\ldots,12$.
Bars report the percentage change in equilibrium lending relative to the no-crisis
counterfactual.}
\label{fig:diate_us}
\end{figure}

\begin{figure}[htbp]
\centering
\includegraphics[width=0.5\textwidth]{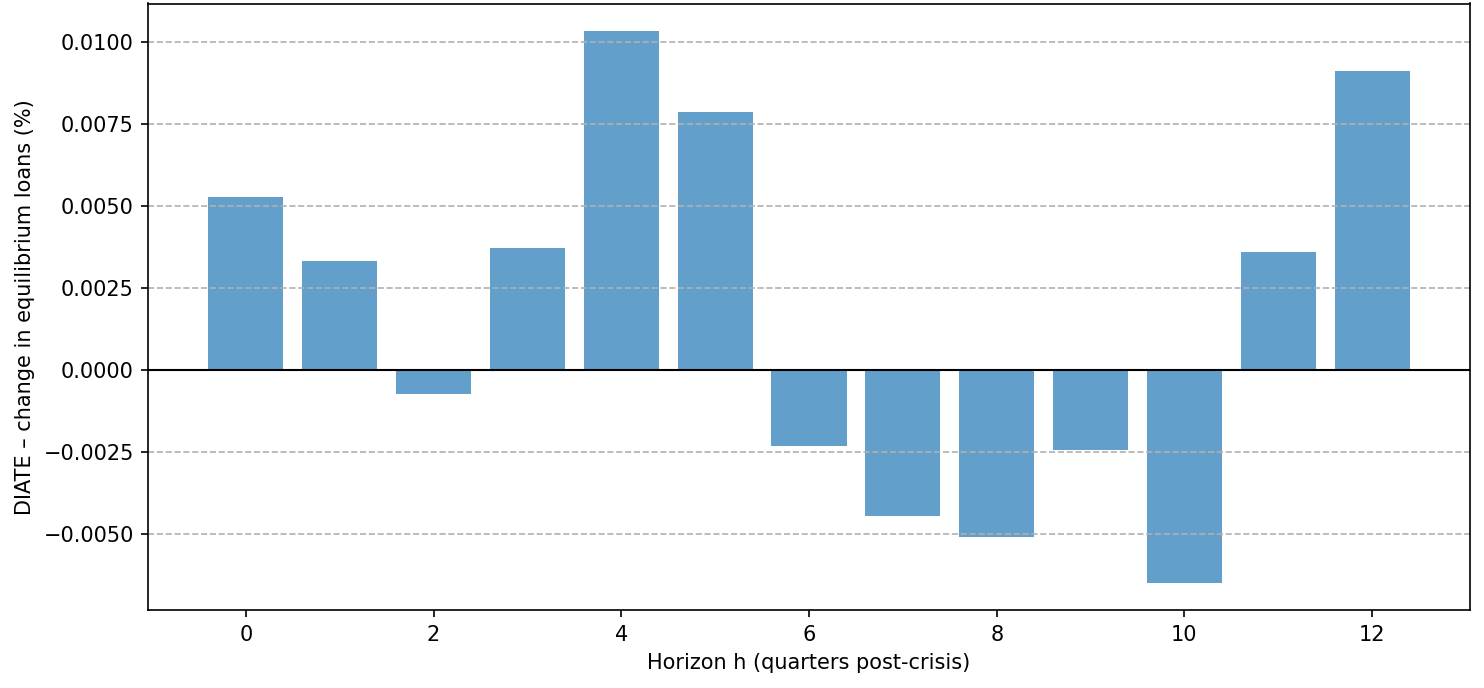}\hfill
\includegraphics[width=0.5\textwidth]{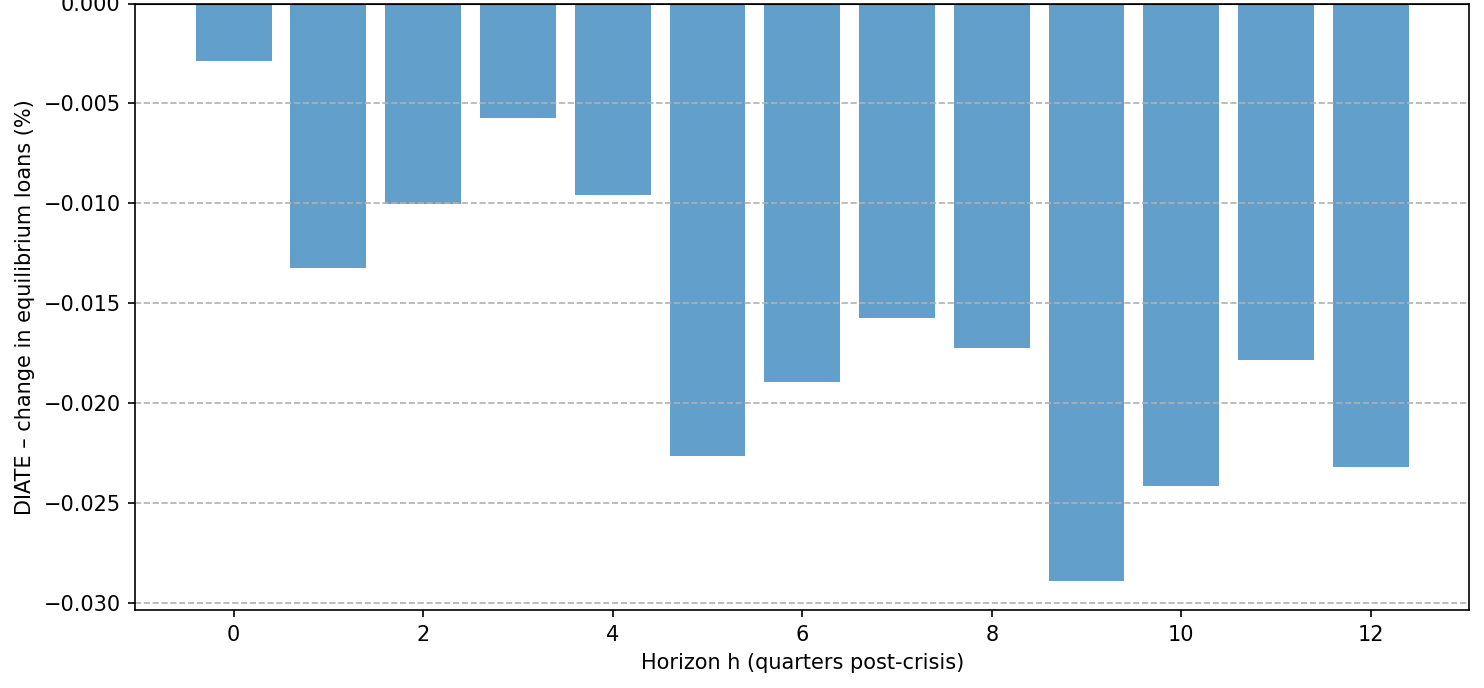}

\parbox{0.48\textwidth}{\small\centering\textit{GFC (2008Q4)}}%
\hfill
\parbox{0.48\textwidth}{\small\centering\textit{Liquidity Crisis (2015Q4)}}

\vspace{1em}

\includegraphics[width=0.48\textwidth]{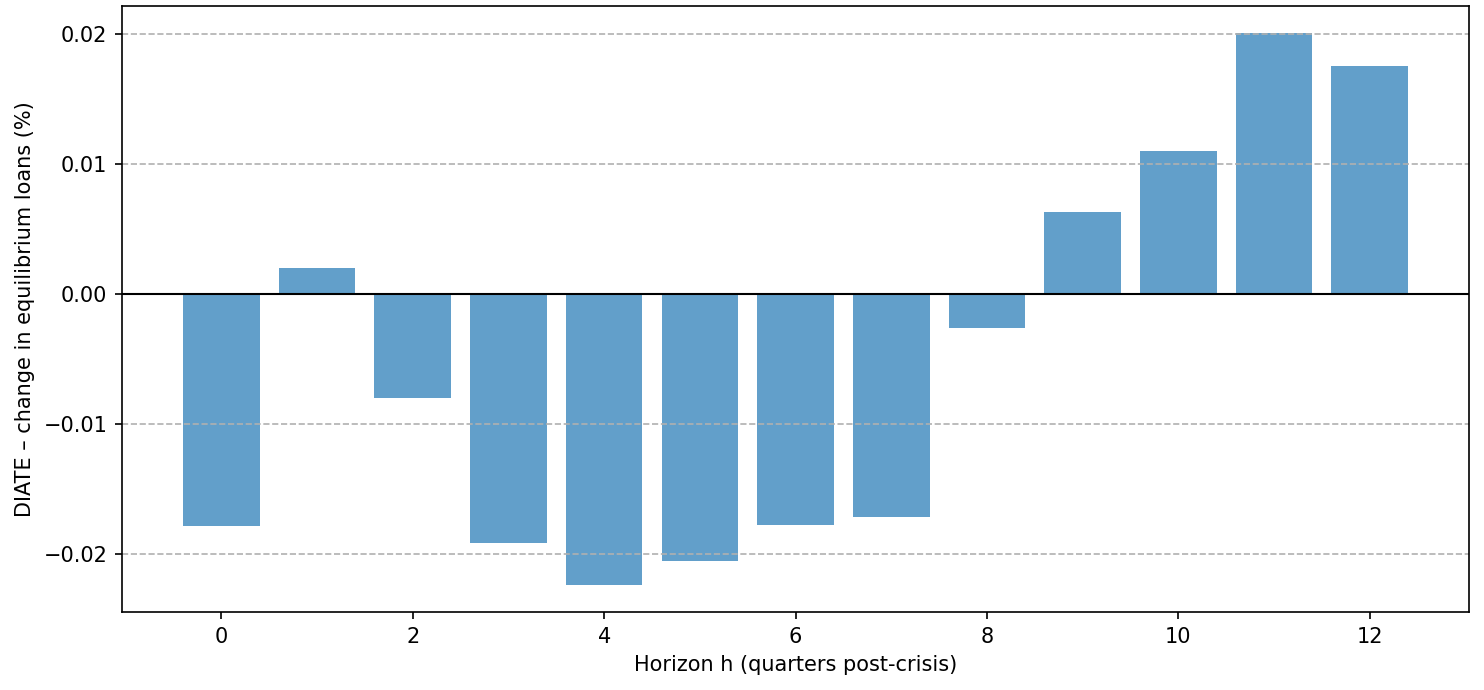}

\parbox{0.48\textwidth}{\small\centering\textit{COVID-19 (2020Q1)}}

\caption{Brazil DIATE Profile. Notes as in Figure~\ref{fig:diate_us}.}
\label{fig:diate_brazil}
\end{figure}

The preceding analysis established that crisis episodes generate persistent changes in intermediary lending capacity. We now quantify the implications of those capacity movements for equilibrium credit supply. Figures~\ref{fig:diate_us}--\ref{fig:diate_brazil} report the DIATE, while Table~\ref{tab:diate} summarizes estimates at selected horizons.

The DIATE maps changes in the latent capacity parameter, $\lambda_{it}$, into changes in equilibrium lending through the model-implied supply constraint. It therefore provides a direct measure of the economic consequences of capacity disruptions beyond the reduced-form dynamics documented in the impulse-response analysis.

\paragraph{United States.}

The estimated lending effects are generally transitory and exhibit substantial mean reversion. During the Global Financial Crisis, lending initially contracts but subsequently rebounds, becoming positive at intermediate horizons and peaking approximately eight quarters after the shock. The pattern is consistent with a temporary disruption followed by balance-sheet repair and policy-supported normalization of intermediary funding conditions.

The post-QE III episode exhibits a modest positive lending response at short horizons that gradually dissipates over time. The COVID-19 episode displays a similar pattern: lending initially benefits from extraordinary policy support but subsequently declines as emergency facilities are withdrawn and funding conditions normalize. Across all three episodes, estimated effects converge toward zero within approximately eight to twelve quarters.

Taken together, the U.S. results suggest that funding-capacity shocks have limited long-run consequences for aggregate lending. Although crisis events temporarily affect intermediary balance sheets, the combination of high baseline capacity and policy intervention prevents persistent contractions in credit supply.

\paragraph{Brazil.}

The Brazilian evidence differs along two dimensions. First, lending responses are more persistent. Second, the recovery process is substantially weaker. Following the 2015 liquidity crisis, lending remains below its counterfactual path throughout the post-event horizon. A similar pattern emerges during the COVID-19 episode, where negative lending effects persist for several years after the initial shock.

The Global Financial Crisis represents a partial exception. Although lending initially recovers following the shock, the longer-horizon estimates indicate that these gains are not fully sustained. Relative to the United States, the Brazilian financial system exhibits a substantially slower reversion toward its pre-crisis lending trajectory.

These findings are consistent with the model's mechanism linking lending outcomes to intermediary funding capacity. Because Brazilian institutions operate with lower baseline values of $\lambda$, a given deterioration in funding conditions pushes a larger fraction of intermediaries toward the binding region of the lending constraint. Consequently, capacity shocks translate more directly into persistent lending contractions.

\paragraph{Cross-country comparison.}

A central implication of the model is that the economic consequences of a crisis depend not only on the magnitude of the shock but also on the level of pre-existing lending capacity. The DIATE estimates provide strong support for this prediction. Despite facing common global shocks during 2008 and 2020, the two countries exhibit markedly different lending responses. The United States experiences relatively short-lived deviations from trend, whereas Brazil displays persistent reductions in equilibrium lending.

The evidence therefore suggests that differences in intermediary funding capacity are an important determinant of credit-market resilience. Economies characterized by deeper funding relationships and larger capacity buffers experience smaller and less persistent lending contractions following adverse shocks.

%

\begin{table}[htbp]
\centering
\caption{Dynamic Average Treatment Effect (DIATE) on Equilibrium Log Lending}
\label{tab:diate}
\begin{threeparttable}
\smallskip

\textsc{Panel A: United States (percent)}
\smallskip

\begin{tabular}{%
  l
  S[table-format=+1.2, retain-explicit-plus]
  S[table-format=+1.2, retain-explicit-plus]
  S[table-format=+1.2, retain-explicit-plus]
  S[table-format=+1.2, retain-explicit-plus]
  S[table-format=+1.2, retain-explicit-plus]
}
\toprule
 & {$h=0$} & {$h=1$} & {$h=4$} & {$h=8$} & {$h=12$} \\
\midrule
GFC (2008Q4)               & -0.33 & -0.94 & +0.18 & +4.92 & +2.67 \\[3pt]
QE\,III\,/\,Liquidity (2014/15) & +1.29 & +1.57 & +0.92 & -0.74 & -1.95 \\[3pt]
COVID-19 (2020Q1)          & +2.19 & +0.49 & -0.70 & -1.68 & -5.94 \\
\bottomrule
\end{tabular}

\bigskip

\textsc{Panel B: Brazil (basis points)}
\smallskip

\begin{tabular}{%
  l
  S[table-format=+1.2, retain-explicit-plus]
  S[table-format=+1.2, retain-explicit-plus]
  S[table-format=+1.2, retain-explicit-plus]
  S[table-format=+1.2, retain-explicit-plus]
  S[table-format=+1.2, retain-explicit-plus]
}
\toprule
 & {$h=0$} & {$h=1$} & {$h=4$} & {$h=8$} & {$h=12$} \\
\midrule
GFC (2008Q4)               & +0.53 & +0.33 & +1.03 & -0.51 & +0.91 \\[3pt]
QE\,III\,/\,Liquidity (2014/15) & -0.29 & -1.33 & -0.96 & -1.72 & -2.32 \\[3pt]
COVID-19 (2020Q1)          & -1.79 & +0.20 & -2.24 & -0.26 & +1.75 \\
\bottomrule
\end{tabular}

\begin{tablenotes}[flushleft]
\footnotesize
\item \textit{Notes.}
  DIATE is defined as
  $\widehat{\mathrm{DIATE}}(h)
    = \overline{\log L^{*}(\hat{\lambda}^{(1)}_{t+h})
      - \log L^{*}(\hat{\lambda}^{(0)}_{t+h})}$,
  averaged over treated$\,\times\,$post observations with $\hat{\mu}_{it}>0$.
  The counterfactual path $\hat{\lambda}^{(0)}$ is taken from the LP-IRF,
  calibrated so the pre-crisis perturbation equals half the estimated crisis impact.
  Brazil values are in basis points; multiply by 100 to convert to percent.
  Full horizon profiles appear in Figures~\ref{fig:diate_us}--\ref{fig:diate_brazil}.
\end{tablenotes}
\end{threeparttable}
\end{table}
\subsection{Persistence and Institutional Heterogeneity}

We next examine the dynamic parameters governing the evolution of lending capacity. Estimation of equation \eqref{eq:panel_ar} yields highly persistent capacity dynamics in both countries. The implied persistence parameters are approximately $\hat{\rho}*{US}=0.93$ and $\hat{\rho}*{BR}=0.90$, corresponding to half-lives of roughly six to nine quarters.

A notable result is the similarity of these estimates across countries. Although the United States and Brazil exhibit markedly different lending responses to crisis shocks, the persistence of capacity itself is nearly identical. The evidence therefore suggests that cross-country differences in lending outcomes arise primarily from differences in the level of steady-state capacity, $\lambda^{SS}$, rather than from differences in the speed at which capacity shocks dissipate. This finding is consistent with Prediction 1 of the model and reinforces the distinction between the magnitude of available lending capacity and the persistence of capacity disturbances.

The estimated capital-capacity elasticity, $\hat{\psi}$, is positive in both countries. Institutions with stronger capital positions maintain higher levels of lending capacity and experience smaller reductions in funding access following adverse shocks. The result supports the model's mechanism linking capitalization to future funding capacity through equation \eqref{eq:lom} and implies that balance-sheet strength contributes directly to intermediary resilience.

The decomposition of lending-capacity dynamics also reveals substantial heterogeneity across institution types. Binding-capacity episodes occur almost exclusively within the mutual sector, while the largest commercial banks rarely operate at the lending constraint. Consequently, the estimated lending effects reported in the previous section are driven primarily by credit unions and cooperative institutions rather than by the major commercial banks themselves.

This pattern is economically important because the large-bank sector remains central to the formation of intermediary lending capacity. Through their provision of wholesale funding and committed credit lines, large commercial banks determine the effective scale of the funding constraint faced by mutual intermediaries. The results therefore suggest that the transmission of funding shocks operates through a network structure in which capacity originates in the wholesale banking sector but affects lending primarily through downstream retail intermediaries.

\begin{figure}[htbp]
\centering
\includegraphics[width=0.5\textwidth]{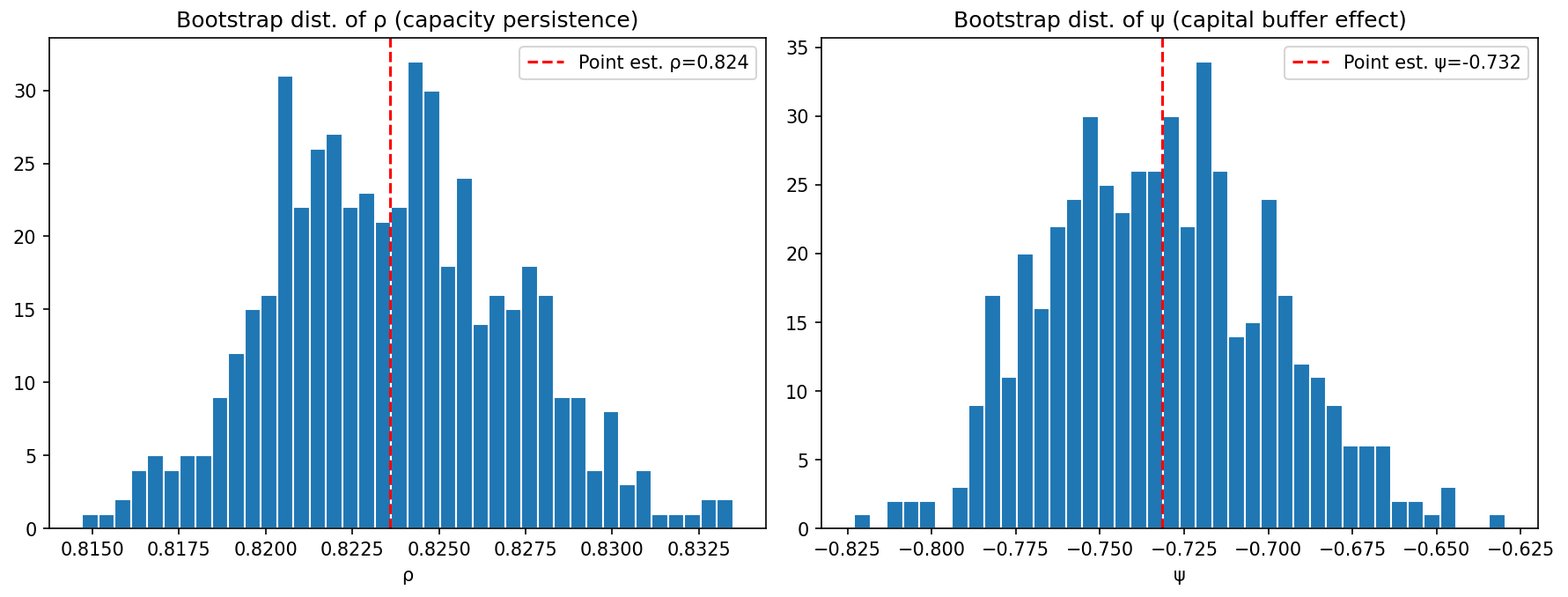}\hfill
\includegraphics[width=0.5\textwidth]{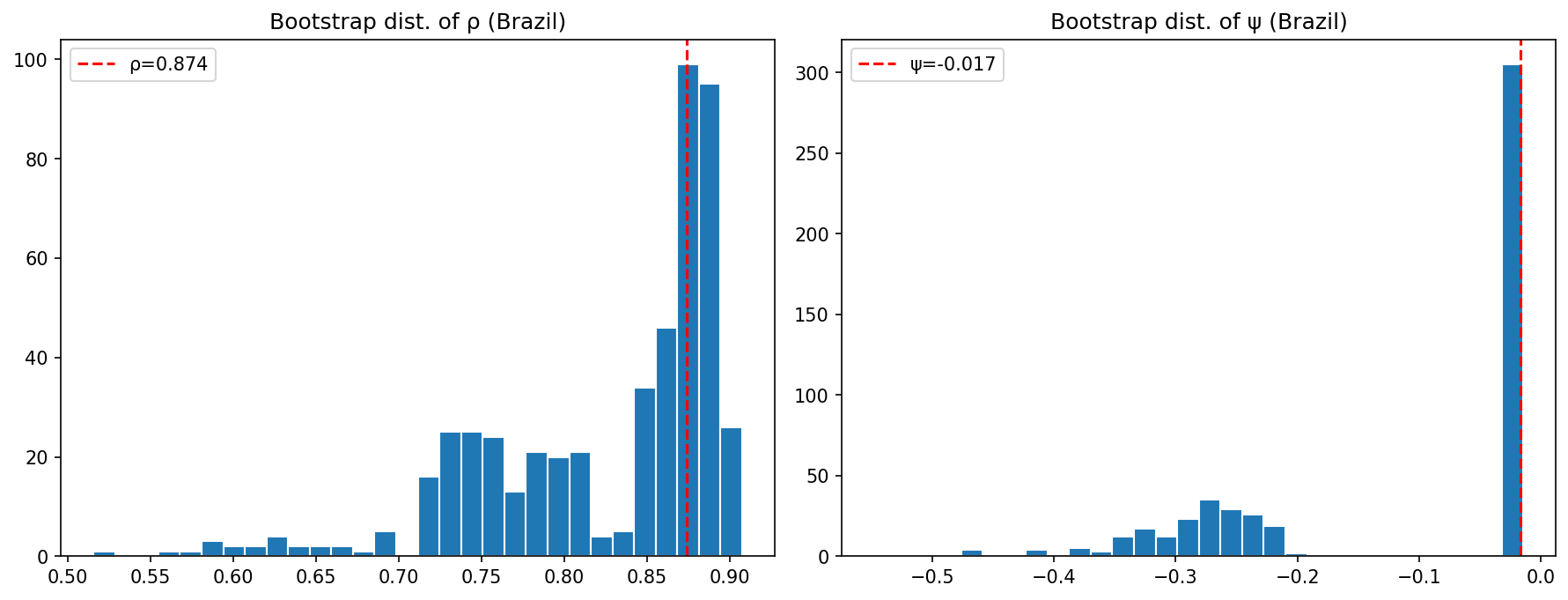}

\vspace{0.5em}
\parbox{\textwidth}{\small\textit{United States \hfill Brazil}}
\caption{Bootstrap Distributions of Dynamic Parameters $(\hat\rho, \hat\psi)$.
Each panel shows histograms from 500 institution-level cluster-bootstrap replications.
The dashed vertical line marks the point estimate from the within-demeaned panel AR(1).}
\label{fig:bootstrap}
\end{figure}

Figure~\ref{fig:bootstrap} reports institution-level cluster-bootstrap distributions for $(\hat{\rho},\hat{\psi})$. The distributions are tightly centered around the point estimates in both countries, although confidence intervals are wider in Brazil owing to the smaller sample size. Overall, the bootstrap evidence indicates that the persistence and capitalization parameters are estimated with sufficient precision to support the dynamic counterfactual exercises that follow.

\subsection{Policy Counterfactuals}

\begin{figure}[htbp]
\centering
\includegraphics[width=0.5\textwidth]{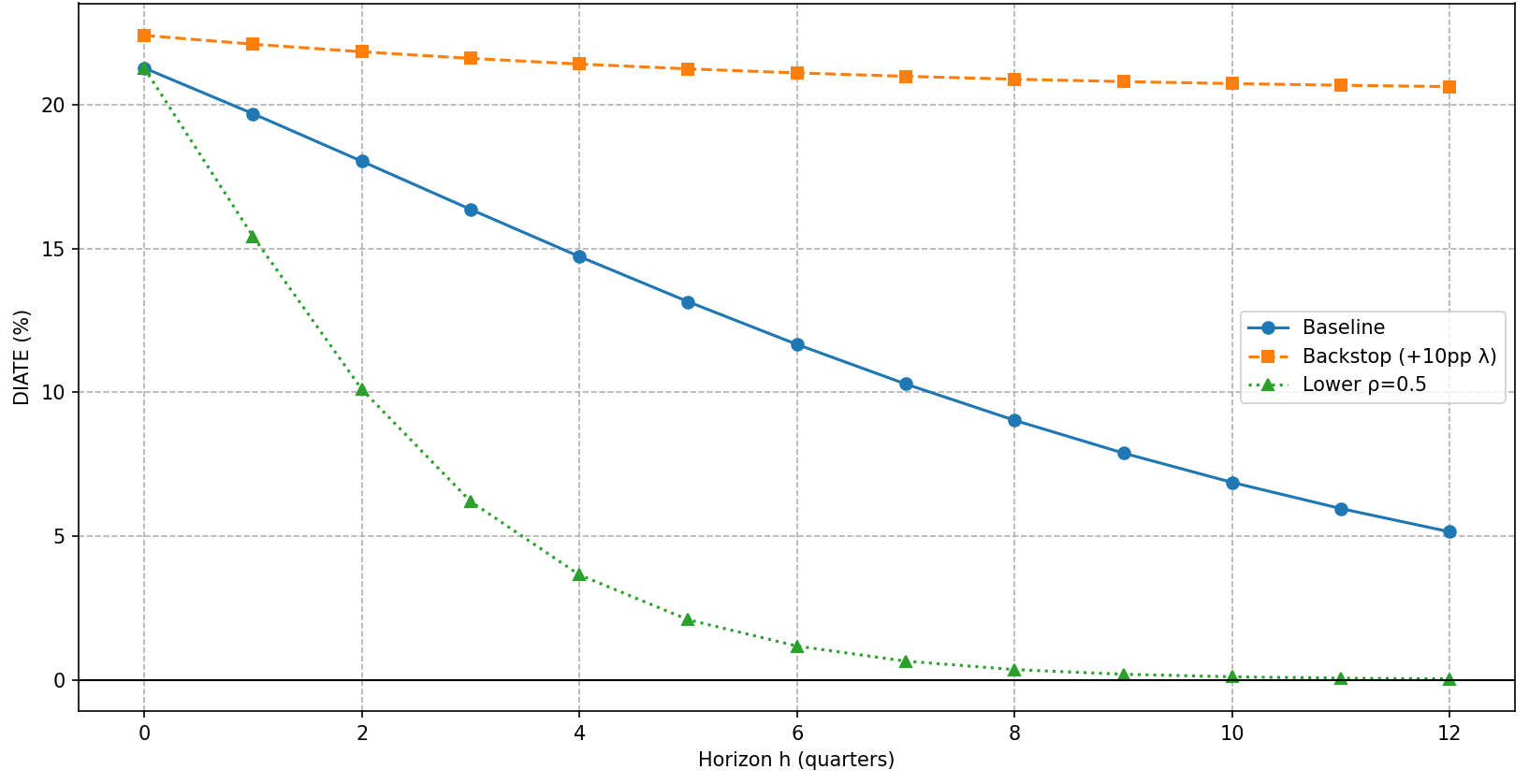}\hfill
\includegraphics[width=0.5\textwidth]{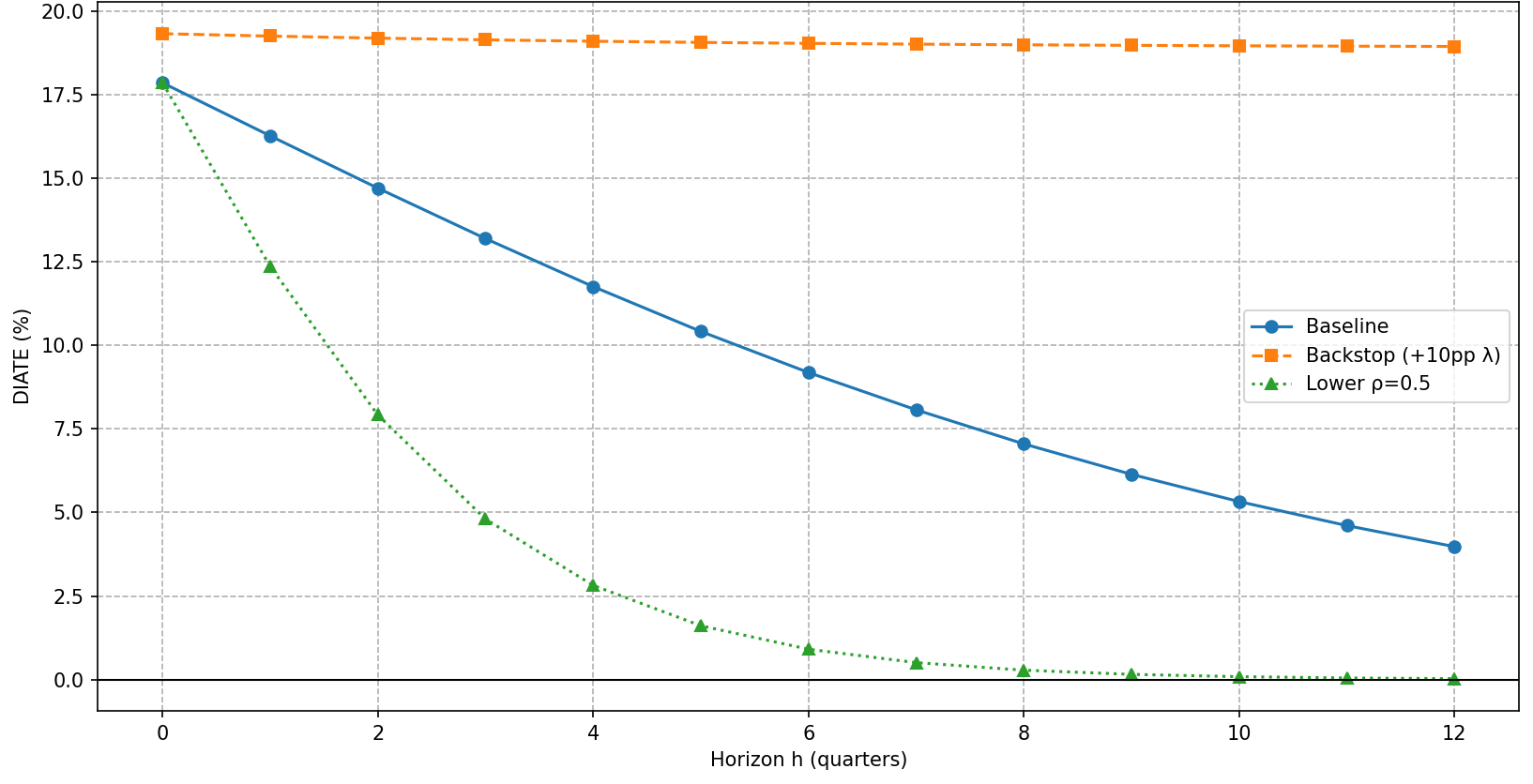}

\parbox{0.48\textwidth}{\small\centering\textit{GFC (2008Q4)}}%
\hfill
\parbox{0.48\textwidth}{\small\centering\textit{End of QE~III (2014Q3)}}

\vspace{1em}

\includegraphics[width=0.5\textwidth]{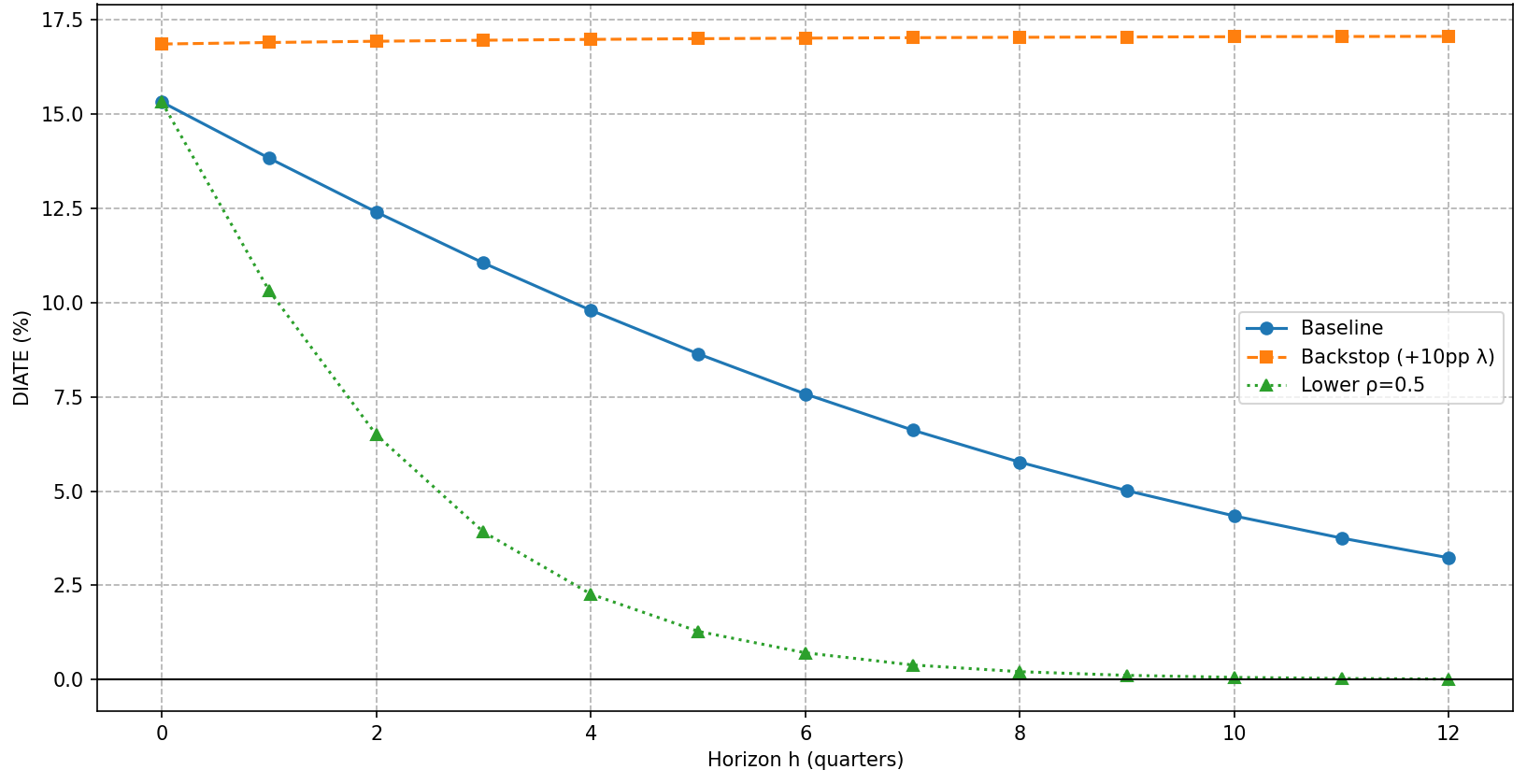}

\parbox{0.48\textwidth}{\small\centering\textit{COVID-19 (2020Q1)}}

\caption{U.S.\ Policy Counterfactual DIATE Paths. Each panel shows the baseline
DIATE (solid), backstop counterfactual (dashed), and alternative-$\rho$ counterfactual
(dotted) from $h=0$ to $h=12$.}
\label{fig:cf_us}
\end{figure}

\begin{figure}[htbp]
\centering
\includegraphics[width=0.5\textwidth]{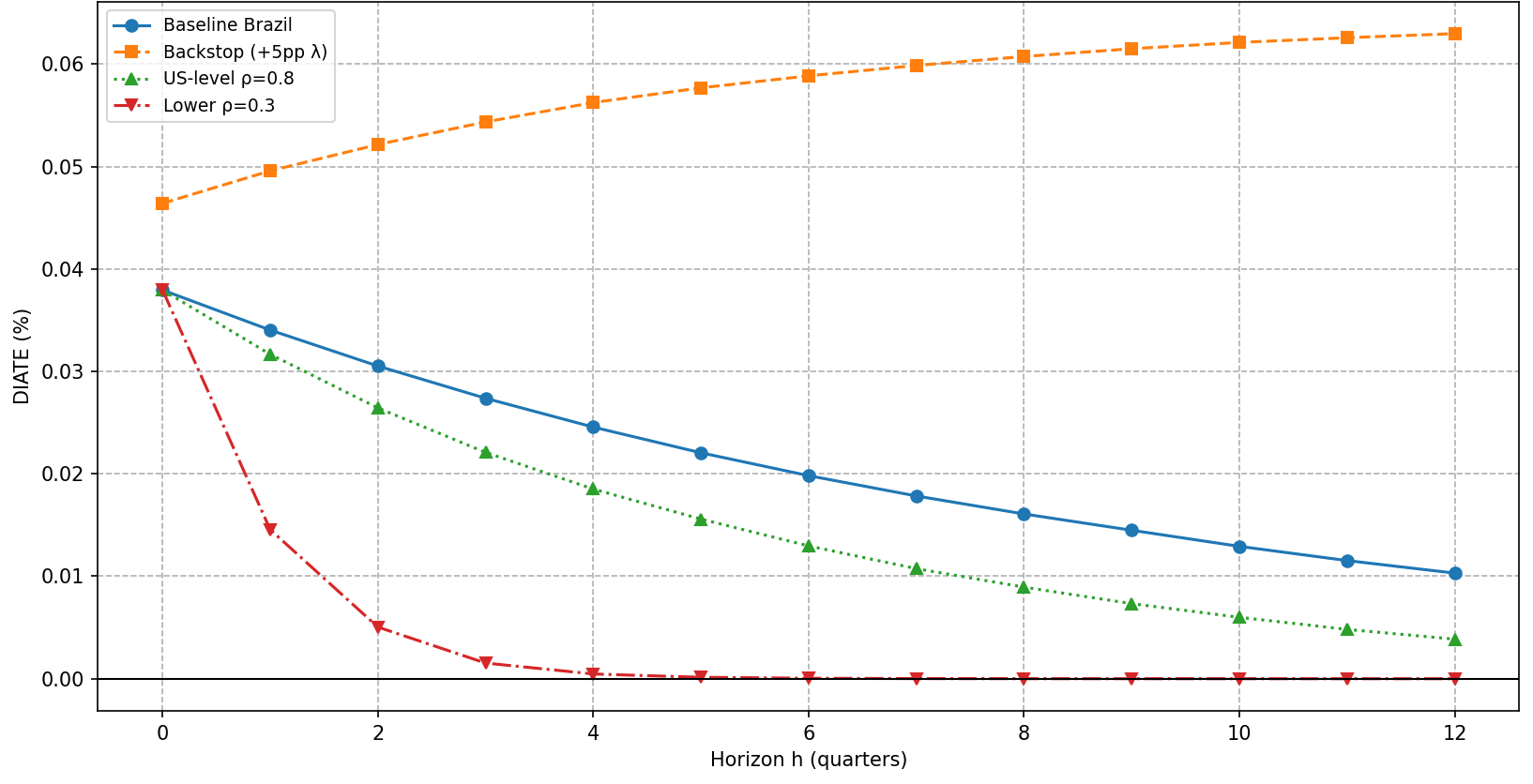}\hfill
\includegraphics[width=0.5\textwidth]{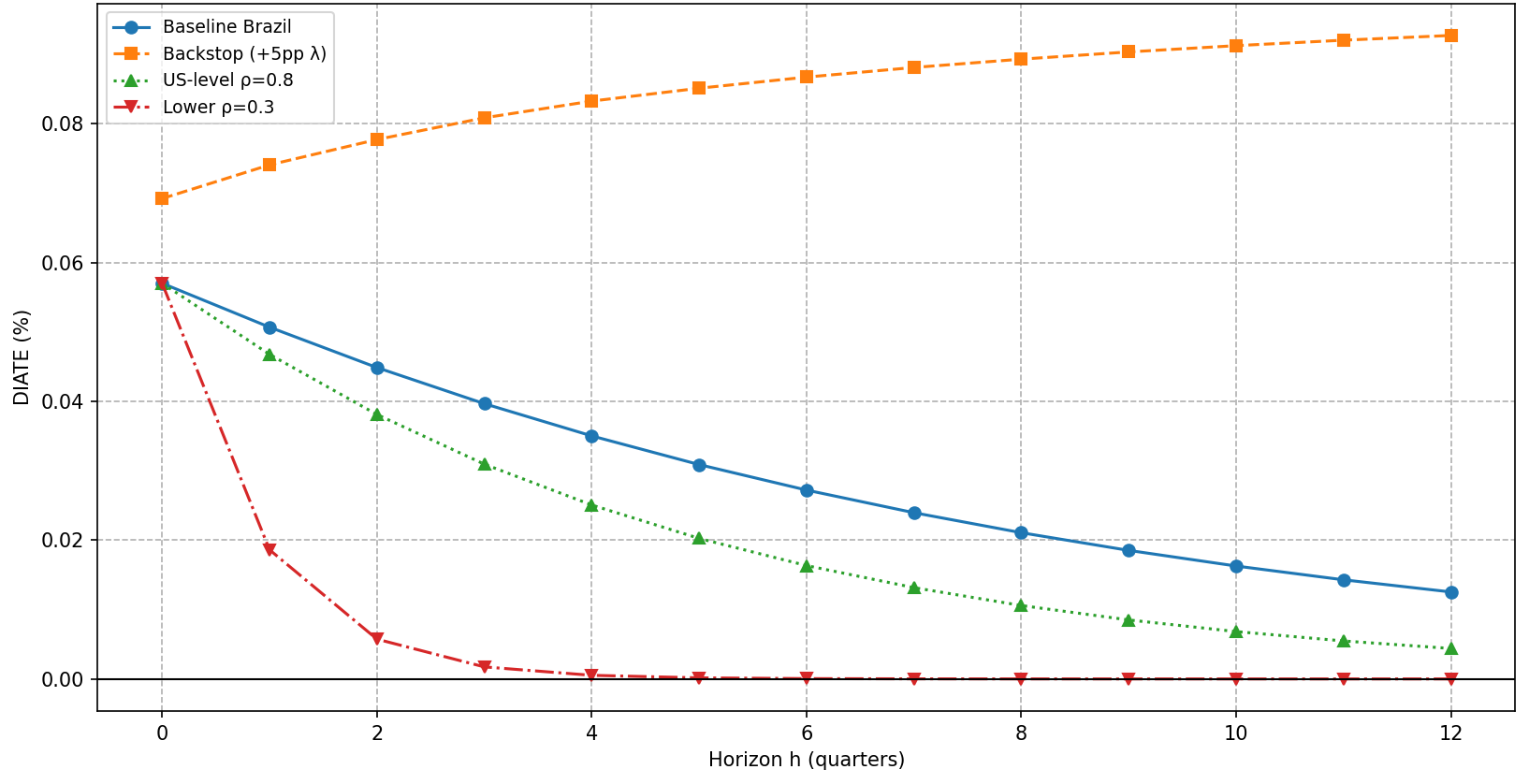}

\parbox{0.48\textwidth}{\small\centering\textit{GFC (2008Q4)}}%
\hfill
\parbox{0.48\textwidth}{\small\centering\textit{Liquidity Crisis (2015Q4)}}

\vspace{1em}

\includegraphics[width=0.5\textwidth]{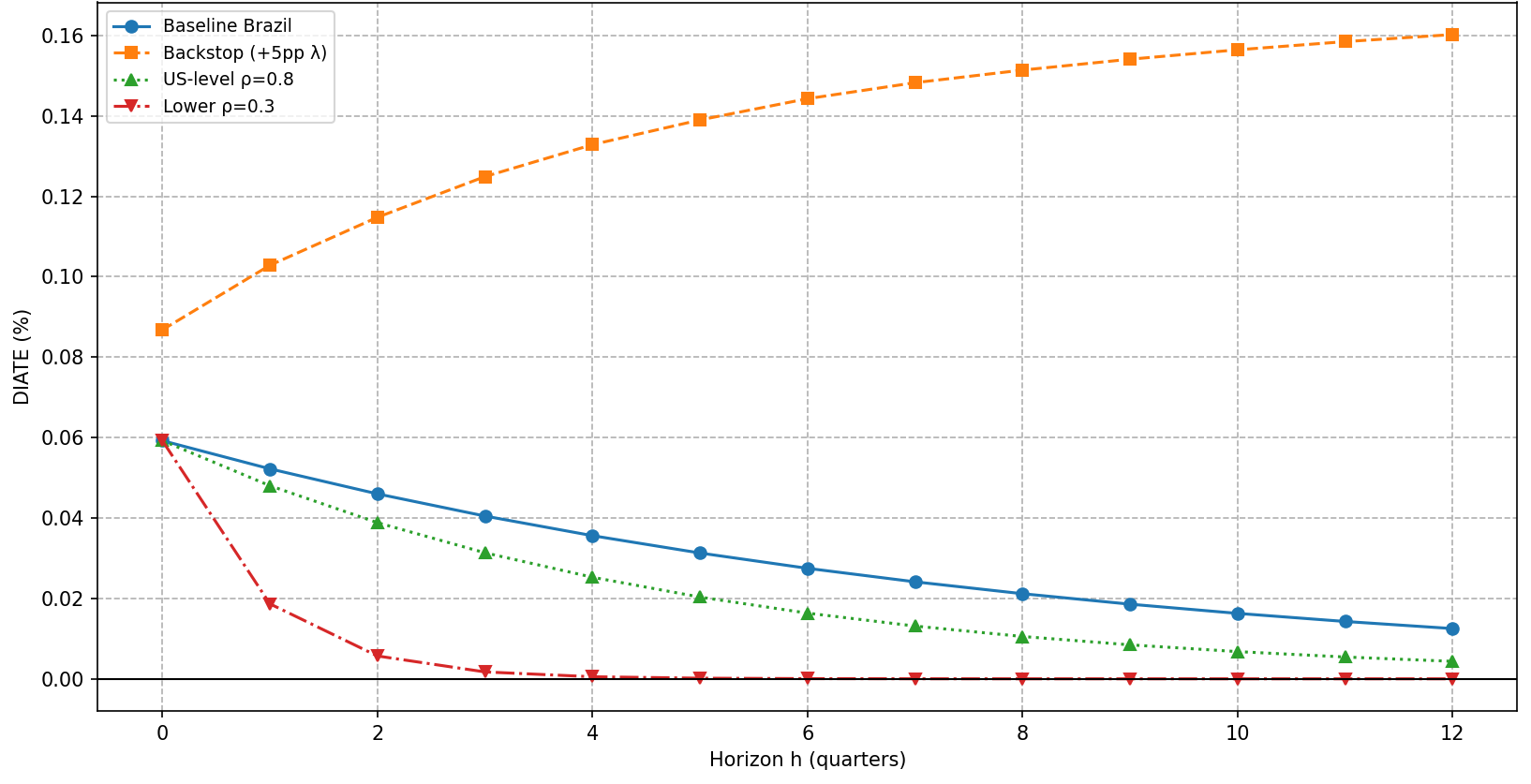}

\parbox{0.48\textwidth}{\small\centering\textit{COVID-19 (2020Q1)}}

\caption{Brazil Policy Counterfactual DIATE Paths. The ``US-$\rho$'' counterfactual
(dotted) shows the capacity path if Brazil's persistence parameter equalled the
U.S.\ proxy.}
\label{fig:cf_brazil}
\end{figure}

We conclude by examining two counterfactual exercises implied by the model. The first introduces a ten-percentage-point increase in lending capacity at the onset of each crisis episode and maintains this intervention throughout the event window. The second replaces each country's estimated persistence parameter with its cross-country counterpart while holding all other parameters fixed. Figures~\ref{fig:cf_us} and \ref{fig:cf_brazil} report the resulting DIATE paths.

\paragraph{Capacity-augmenting interventions.}

The first counterfactual evaluates the effects of policies that directly expand intermediary lending capacity. In the model, such interventions operate through an increase in $\lambda_{it}$ and therefore relax the effective lending constraint. Because capacity enters dynamically through the persistence channel, the impact of the intervention accumulates over time according to the policy multiplier

\[
M^\Xi(h) = 
\gamma_\Xi
\sum_{j=0}^{h}
\hat{\rho}^{,h-j}.
\]

The effects differ substantially across countries. In the United States, the additional capacity generates relatively modest increases in equilibrium lending. Although lending rises following the intervention, the quantitative effects remain limited throughout the horizon. This outcome reflects the fact that only a small fraction of institutions operate near the binding region of the lending constraint, implying that additional capacity is largely inframarginal.

In Brazil, the same intervention produces substantially larger effects. The counterfactual lending response exceeds the baseline path by a wide margin and the gap increases with the forecast horizon. Because lending capacity is considerably lower in steady state, a larger fraction of institutions are capacity constrained. As a result, an identical increase in $\lambda_{it}$ relaxes the binding constraint for a much larger share of intermediaries and generates correspondingly larger lending effects.

These results provide support for Prediction 2 of the model: the effectiveness of capacity-enhancing interventions is increasing in the prevalence of binding constraints. Policies that expand intermediary funding capacity therefore have the greatest effects in environments where lending capacity is initially scarce.

\paragraph{The persistence counterfactual.}

The second exercise isolates the role of capacity persistence. Specifically, we replace each country's estimated persistence parameter with the corresponding estimate from the other country while holding the remaining parameters fixed.

The results indicate that differences in persistence play a comparatively limited role in explaining the cross-country divergence documented above. Replacing Brazil's persistence parameter with the U.S. estimate modestly alters the timing of recovery but leaves the overall magnitude of lending losses largely unchanged. Similarly, imposing Brazil's persistence parameter on the United States has only a limited effect on the qualitative response of lending.

These findings reinforce a central implication of the model. Although persistence governs the speed with which capacity shocks dissipate, long-run lending resilience depends primarily on the level of steady-state capacity. Economies characterized by larger values of $\lambda^{SS}$ experience smaller lending contractions even when shocks are highly persistent, whereas economies with limited baseline capacity remain vulnerable despite relatively similar rates of mean reversion.

Taken together, the counterfactual exercises suggest that cross-country differences in lending outcomes are driven primarily by disparities in intermediary funding capacity rather than by differences in the persistence of capacity shocks. Expanding the effective capacity of constrained intermediaries therefore generates substantially larger gains than altering the speed at which capacity adjusts over time.

\section{Conclusion}
\label{sec:conclusion}

This paper develops and estimates a dynamic model of intermediary lending capacity in which the link between wholesale funding and credit supply is governed by an endogenous state variable, $\lambda_{it}$. The model highlights a simple but important mechanism: the effects of funding shocks depend not only on the shocks themselves, but also on the amount of lending capacity available before the shock occurs and on the persistence with which capacity evolves over time.

Using supervisory data from the United States and Brazil, we recover institution-level measures of lending capacity and estimate their dynamic response to major crisis episodes. Three findings emerge. First, lending capacity is substantially higher in the United States than in Brazil, with estimated steady-state capacity differing by a factor of roughly three to six across episodes. Second, capacity dynamics are highly persistent in both countries, implying similar half-lives of capacity shocks. Third, despite similar persistence, lending responses differ markedly across countries because institutions operate with very different levels of baseline capacity. Economies characterized by higher lending capacity experience smaller and less persistent contractions in credit supply following adverse funding shocks.

The counterfactual analysis reinforces this conclusion. Capacity-enhancing interventions generate their largest effects in environments where a substantial fraction of intermediaries operate near the lending constraint. By contrast, differences in the persistence of capacity shocks account for relatively little of the observed cross-country variation in lending outcomes. The evidence therefore points to the level of lending capacity, rather than its rate of adjustment, as the primary determinant of credit-market resilience.

More broadly, the results suggest that the structure of intermediary funding relationships plays a central role in the transmission of financial shocks. Existing research has emphasized capital adequacy, liquidity management, and regulatory constraints as determinants of financial stability. The evidence presented here highlights an additional margin: the effective capacity with which wholesale funding can be transformed into credit supply. Understanding how this capacity is accumulated, maintained, and supported during periods of stress remains an important direction for future research.

\newpage
\bibliographystyle{apalike}
\bibliography{References}

\newpage
\appendix
\section{Dynamic Recursive Model: Complete Derivations}
\label{app:model}

This appendix presents the complete derivations of the dynamic recursive model
referenced in Section~\ref{sec:model}. All notation is as defined in the main text.

\subsection{Environment and State Space}

Consider a continuum of mutual intermediaries $i\in[0,1]$ operating in discrete
time $t=0,1,2,\ldots$. The state vector is
\begin{equation}
s_{it} = (K_{it},\;D_{it},\;BL_{it},\;\lambda_{it},\;\om_{it},\;Z_t),
\end{equation}
where $K_{it}$ is equity capital (capital-to-asset ratio $k_{it}\equiv K_{it}/A_{it}$),
$D_{it}$ and $BL_{it}$ are deposits and bank credit lines (both predetermined from
the intermediary's perspective), $\lambda_{it}$ is the endogenous capacity multiplier,
$\om_{it}$ is the funding cost from \eqref{eq:supply}, and $Z_t=(C_t,\Xi_t,i_t,
\mathrm{Risk}_t)$ is the exogenous aggregate state.

\subsection{Bellman Equation and Period Profit}

From inverting household loan demand \eqref{eq:demand}, the period profit is
\begin{equation}
\Pi_{it}(L_{it}) = a_{it}L_{it} - \frac{b}{2}L_{it}^2,\quad
a_{it} = \frac{A_{jt}+\zeta Z_t}{\chi} - \om_{it},\quad
b = \frac{1+\chi\phi}{\chi}.
\end{equation}
The Bellman equation is
\begin{equation}
V(s_{it}) = \max_{0\le L\le D+\lambda BL}\left\{\Pi(L)+\beta\,\mathbb{E}_t[V(s_{i,t+1})]\right\}.
\end{equation}

\subsection{Optimal Period Policy (KKT)}

The Lagrangian with multiplier $\mu_{it}\ge 0$ on the capacity ceiling is
\begin{equation}
\mathcal{L} = a_{it}L - \frac{b}{2}L^2 + \mu_{it}(D_{it}+\lambda_{it}BL_{it}-L)
+ \beta\,\mathbb{E}_t[V(s_{i,t+1})].
\end{equation}
The FOC $a_{it}-bL-\mu_{it}=0$ with complementary slackness yields:
\begin{align}
L^U_{it} &= \frac{a_{it}}{b} = \frac{A_{jt}+\zeta Z_t - \chi\om_{it}}{1+\chi\phi},\quad
\mu_{it}=0 \text{ (slack)},\\
L^*_{it} &= D_{it}+\lambda_{it}BL_{it},\quad
\mu^*_{it} = \frac{1+\chi\phi}{\chi}(L^U_{it}-D_{it}-\lambda_{it}BL_{it})>0\text{ (binding)},
\end{align}
giving the unified equilibrium $L^*_{it}=\min\{L^U_{it},D_{it}+\lambda_{it}BL_{it}\}$.

\subsection{Laws of Motion}

Capital accumulates as
$K_{i,t+1}=(1-\delta_K)K_{it}+\Pi^*_{it}-F$
where $\delta_K$ is the depreciation/dividend rate and $F$ is a fixed operating cost.
The capacity law of motion is
\begin{equation}
\lambda_{i,t+1} = \rho\,\lambda_{it} + (1-\rho)\left[\bar\lambda_t + \psi k_{it}\right]
+ \sigma_\lambda\xi_{it},\quad \xi_{it}\sim\mathcal{N}(0,1)\;\text{i.i.d.},
\end{equation}
where the aggregate target $\bar\lambda_t=\bar\lambda+\alpha_\lambda C_t+\gamma_\Xi\Xi_t$.
A crisis shock $C_t=1$ produces an immediate contraction $\alpha_\lambda<0$.

\subsection{Envelope Conditions and Dynamic Shadow Price}

Applying the Envelope Theorem to the Bellman equation with respect to $\lambda_{it}$:
\begin{equation}
\frac{\partial V}{\partial\lambda_{it}} = \mu^*_{it}BL_{it}
+ \beta\rho\,\mathbb{E}_t\!\left[\frac{\partial V(s_{i,t+1})}{\partial\lambda_{i,t+1}}\right].
\end{equation}
Defining the dynamic shadow price $p_t\equiv\partial V/\partial\lambda_{it}$:
\begin{equation}
p_t = \mu^*_{it}BL_{it} + \beta\rho\,\mathbb{E}_t[p_{t+1}]
= \sum_{h=0}^\infty(\beta\rho)^h\,\mathbb{E}_t\!\left[\mu^*_{i,t+h}BL_{i,t+h}\right].
\end{equation}

\begin{proposition}[Dynamic Shadow Value]
\label{prop:shadow}
The dynamic shadow value of capacity is the discounted sum of all future
constraint-binding shadow costs at discount rate $\beta\rho$. It exceeds the static
shadow value whenever $\rho>0$ and the binding probability is positive in future periods.
\end{proposition}

\subsection{Steady State}

In the no-crisis steady state ($C_t=0$, $\Xi_t=\Xi_0$), the cross-sectional
distribution of $\lambda$ is:
\begin{equation}
\lambda^{SS}_i \sim \mathcal{N}\!\left(\bar\lambda+\gamma_\Xi\Xi_0+\psi k^{SS},\;
\frac{\sigma_\lambda^2}{1-\rho^2}\right).
\end{equation}

\begin{proposition}[Steady-State Capacity Dispersion]
\label{prop:dispersion}
Cross-sectional dispersion in capacity is increasing in $\rho$ and $\sigma_\lambda$.
Institutions with higher steady-state capital ratio $k^{SS}$ operate at strictly higher
steady-state capacity. A policy backstop $\Xi_0$ raises aggregate steady-state capacity
by $\gamma_\Xi$ per unit.
\end{proposition}

\subsection{Analytical Impulse Response and Half-Life}

Starting from steady state $\lambda^{SS}$, a crisis shock at $t=0$ produces the
conditional capacity path:
\begin{equation}
\mathbb{E}_0[\lambda_{i,h}] = \lambda^{SS} + \rho^h\alpha_\lambda + (1-\rho^h)\gamma_\Xi
\mathbb{E}_0\!\left[\sum_{j=1}^h\rho^{h-j}(\Xi_j-\Xi_0)\right].
\end{equation}
Absent policy intervention:
\begin{equation}
\mathrm{IRF}_\lambda(h) = \rho^h\alpha_\lambda,
\end{equation}
and the half-life of the contraction is
\begin{equation}
h_{1/2} = \frac{\log 2}{\log(1/\rho)}.
\end{equation}

\begin{corollary}[Half-Life]
\label{cor:halflife}
For $\rho=0.93$ (U.S.): $h_{1/2}\approx 9.5$ quarters. For $\rho=0.90$ (Brazil):
$h_{1/2}\approx 6.6$ quarters. The similar persistence means the difference in
recovery speed across countries is driven primarily by the level of $\lambda^{SS}$,
not by asymmetric persistence.
\end{corollary}

\subsection{DIATE and Decomposition}

\begin{definition}[Dynamic Average Treatment Effect]
\begin{equation}
\mathrm{DIATE}(h) = \mathbb{E}\!\left[\log L^*\!\left(\lambda^{(1)}_{t+h}\right)
- \log L^*\!\left(\lambda^{(0)}_{t+h}\right)\;\middle|\; C_t=1,\;\hat\mu_{it}>0\right].
\end{equation}
\end{definition}

\begin{proposition}[Dynamic Decomposition]
\label{prop:decomp}
$\mathrm{DIATE}(h) = \mathrm{DIATE}^{\mathrm{cost}}(h)+\mathrm{DIATE}^{\mathrm{constr}}(h)$,
where the constraint channel is:
\begin{equation}
\mathrm{DIATE}^{\mathrm{constr}}(h) \approx
\frac{\mathrm{IRF}_\lambda(h)\cdot\mathbb{E}[BL_{t+h}]}
{\mathbb{E}[D_{t+h}+\lambda^{(1)}_{t+h}BL_{t+h}]},
\end{equation}
and the cost channel is:
\begin{equation}
\mathrm{DIATE}^{\mathrm{cost}}(h) \approx
-\frac{\theta_2\,\mathrm{IRF}_{r^{BL}}(h)}{\phi}\cdot
\mathbb{E}\!\left[\frac{\partial\log L^U}{\partial\om}\right].
\end{equation}
\end{proposition}

\subsection{Policy Backstop Multiplier}

A backstop $\Xi^*>\Xi_0$ activated at $t=0$ for $T$ periods shifts the capacity path:
\begin{equation}
\lambda^{\Xi^*}_{i,t+h} = \lambda^{(1)}_{i,t+h}
+ \gamma_\Xi(\Xi^*-\Xi_0)\sum_{j=0}^{\min(h,T)}\rho^{h-j}.
\end{equation}
The implied policy multiplier on equilibrium loans is:
\begin{equation}
M^\Xi(h) = \gamma_\Xi\cdot\frac{\partial\mathrm{DIATE}(h)}{\partial\lambda^{(1)}_{t+h}}
\cdot\sum_{j=0}^{\min(h,T)}\rho^{h-j},
\label{eq:mult}
\end{equation}
which is increasing in $\gamma_\Xi$, $\rho$, and the binding share. The backstop is
thus most powerful in thin-capacity systems (low $\lambda^{SS}$, many binding
institutions) with high persistence ($\rho$) -- precisely the profile of the Brazilian
system relative to the U.S.

\subsection{Cross-Country Counterfactual}

The Brazil-with-US-persistence counterfactual capacity path is:
\begin{equation}
\lambda^{BR\to US}_{i,t+h} = \lambda^{SS}_{BR} + \rho_{US}^h\,\alpha_{\lambda,BR},
\end{equation}
where $\rho_{US}=0.80$ is used as the proxy for U.S.-level persistence in the
Brazilian counterfactual. Since $\rho_{US}<\hat\rho_{BR}\approx 0.90$, this
counterfactual shows faster mean-reversion from the same initial shock -- corresponding
to the scenario where Brazilian wholesale markets become shallower (shorter-duration
credit lines) but institutional shock absorption at each quarter is larger.

\subsection{Comparative Statics}

\begin{proposition}[Persistence and DIATE Profile]
\label{prop:rho}
Higher $\rho$ raises $|\mathrm{DIATE}(h)|$ for all $h\ge 1$ and increases the
cumulative $\overline{\mathrm{DIATE}}(H)$ for any finite $H$. It has no effect
on the impact DIATE at $h=0$.
\end{proposition}

\begin{proposition}[Capital Buffer and Steady State]
\label{prop:psi}
Higher $\psi$ raises $\lambda^{SS}$ without changing $\rho$, shifting the entire
DIATE profile upward proportionally when the majority of institutions are in the
slack regime.
\end{proposition}

Table~\ref{tab:cstat} summarizes the comparative statics and their empirical mapping.

\begin{table}[htbp]
\centering
\begin{tabular}{lcccc}
\toprule
Parameter & U.S. & Brazil & Direction & Mechanism \\
\midrule
$\rho$ (persistence) & $\approx 0.93$ & $\approx 0.90$ & $\rho_{US}>\rho_{BR}$ & U.S.\ constraints more sluggish \\
$\psi$ (capital premium) & high & low & $\psi_{US}>\psi_{BR}$ & deeper capital markets \\
$\alpha_\lambda$ (crisis impact) & moderate & large & $|\alpha_{BR}|>|\alpha_{US}|$ & Brazil more fragile on impact \\
$\gamma_\Xi$ (policy elasticity) & high & moderate & $\gamma_{US}>\gamma_{BR}$ & Fed facilities more effective \\
$\phi$ (supply slope) & $\approx 0$ & $<0$ & & elastic vs.\ rationing supply \\
$\lambda^{SS}$ (steady state) & 0.45--0.63 & 0.10--0.18 & $3.5\times$ higher in U.S. & institutional depth \\
\bottomrule
\end{tabular}
\caption{Comparative Statics Summary: Predicted Parameter Patterns and Mechanisms.}
\label{tab:cstat}
\end{table}

\subsection{Identification of Dynamic Parameters: Four-Stage Summary}

\textbf{Stage 1.} Static parameters $(\chi,\phi,\theta_1,\theta_2,\theta_3,\gamma_0,
\gamma_1,\gamma_2)$ from demand LIML + supply IV + funding-cost 2SLS as in
Section~\ref{sec:identification}.

\textbf{Stage 2.} Cross-sectional recovery of $\hat\lambda_{it}=(L_{it}-D_{it})/BL_{it}$
for $\hat\mu_{it}>0$; two-way within-demean; winsorize at 1st/99th percentiles.

\textbf{Stage 3.} Panel AR(1) $\hat\lambda_{i,t+1}=\rho\hat\lambda_{it}+\psi k_{it}
+\alpha_i+\tau_t+\varepsilon_{it}$ by within-demeaned OLS with institution-cluster
standard errors. Aggregate parameters $(\alpha_\lambda,\gamma_\Xi)$ from time-series
regression of cross-sectional median $\tilde\lambda_t$ on $C_t$ and $\Xi_t$.

\textbf{Stage 4.} Local projections~\eqref{eq:lp} for the nonparametric IRF;
DIATE~\eqref{eq:diate} at each horizon using the LP-IRF capacity path; 500-replication
institution-level cluster bootstrap for joint confidence intervals.

\end{document}